\documentclass[11pt,a4paper,onecolumn]{IEEEtran}
\def\IncludeSupplementary{true} 

\usepackage[left=2cm,right=2cm,top=2cm,bottom=2cm]{geometry}
\usepackage[dvipsnames]{xcolor}
\usepackage[utf8]{inputenc}  
\usepackage{url}
\usepackage{amsmath}
\usepackage[makeroom]{cancel}
\usepackage{amsfonts}
\usepackage{amssymb}
\usepackage[pdftex]{graphicx}
\graphicspath{{../pdf/}{../jpeg/}}
\usepackage{pgfplots}
\pgfplotsset{ 
  compat=newest, 
}
\usepgfplotslibrary{fillbetween}
\usetikzlibrary {positioning}
\DeclareGraphicsExtensions{.pdf,.jpeg,.png}
\usepackage{tikz}

\author{Emma F.~Thomas$^{1,\star}$, Mengbin Ye$^2$, Simon D.~Angus$^3$, Tony J.~Mathew$^2$, Winnifred Louis$^4$, \\Liam A. Walsh$^2$, Silas Ellery$^1$, Morgana Lizzio-Wilson$^5$, and Craig McGarty$^6$.\\ \vspace{11pt}
\small{$^1$College of Education, Psychology and Social Work, Flinders University, Adelaide, Australia\\
$^2$Centre for Optimisation and Decision Science, Curtin University, Perth, Australia\\
$^3$Department of Economics \& SoDa Laboratories, Monash University, Melbourne, Australia \\
$^4$School of Psychology, University of Queensland, Brisbane, Australia\\
$^5$Department of Psychology, University of Exeter, Exeter, United Kingdom\\
$^6$School of Social Sciences and Psychology, Western Sydney University, Sydney, Australia\\
$^\star$Corresponding author: e-mail: emma.thomas@flinders.edu.au  }
}
\title{\LARGE {Repeated and incontrovertible collective action failure leads to protester disengagement and radicalisation }}
\usepackage{tabularx}
\usepackage{float}
\usepackage{amsthm}  
\usepackage{enumitem} 
\usepackage{bbm} 
\usepackage{cite}
\usepackage{subfig}
\usepackage{comment}
\usepackage{booktabs}
\usepackage{marginnote}
\usepackage[pagewise]{lineno}
\usepackage[normalem]{ulem} 

\renewcommand{\eqref}[1]{Eq.~(\ref{#1})}  

\newcommand{\vertiii}[1]{{\left\vert\kern-0.25ex\left\vert\kern-0.25ex\left\vert #1 
    \right\vert\kern-0.25ex\right\vert\kern-0.25ex\right\vert}} 

\date{}

\newcommand{\tm}[1]{{\color{orange}#1}}

\newcommand{\deleted}[1]{} 

\usepackage{hyperref} 
\hypersetup{
	citebordercolor = 1 1 1,
	linkbordercolor = 1 1 1,
	menubordercolor = 1 1 1,
}

\usepackage{cleveref}

\begin{document}
\maketitle

\begin{abstract}
Protest is ubiquitous in the 21st Century and the people who participate in such movements do so because they seek to bring about social change. However, social change takes time and involves repeated interactions between individual protesters, social movements and the authorities to whom they appeal for change. These complexities of time and scale have frustrated efforts to isolate the conditions that foster an enduring movement, on the one hand, and the adoption of more radical (unconventional, unacceptable) tactics on the other. Here, we present a novel, theoretically informed and empirically evidenced, agent-based model of collective action that provides a unified framework to address these dual challenges. We model ~10,000 iterations within a simulated society and show that where an authority is responsive, and protesters can (cognitively and/or socially) contest the failure of their movement, a moderate conventional movement prevails. Conversely, where an authority repeatedly and incontrovertibly fails the movement, the population disengages but becomes radicalised (latent radicalism). This latter finding, whereby the whole population is disengaged but prepared to use radical methods to bring about social change, likely reflects the febrile pre-cursor state to sudden, revolutionary change. Results highlight the potential for simulations to reveal emergent, as-yet under-theorized, phenomena.
\vspace{22pt}
\end{abstract}

We are living in an age of protest\cite{ortiz2022analysis,cantoni2024protests}. From the Arab Spring Revolutions in the Middle East and Northern Africa, the toppling of longstanding dictatorships in Sudan and Algeria, the Yellow Vests movement in France, pro-democracy movements in Chile, Venezuela, Thailand, Hong Kong, as well as the Black Lives Matter and MeToo movements, such movements have played a critical role in countering authoritarian rule, advancing the rights and freedoms of people across the world \cite{roggeband2017handbook, shuman2023social}. Mirroring the real world proliferation of such movements, there are flourishing interdisciplinary literatures on protest~\cite{cantoni2024protests,paret2021persistent,saunders2012explaining}, social movement participation~\cite{tilly2015social,mccarthy1977resource,mcadam1999political,della2015oxford} and collective action~\cite{agostini2021toward,becker2017dynamic,thomas2022mobilise,tarrow1993cycles}. These literatures tell us much about why people engage with such movements: hundreds of studies confirm that collective action emerges among people who are morally convicted about their stance, feel the status quo is unjust, believe that by acting together they can change the situation, and identify with groups that can mobilise action \cite{agostini2021toward,becker2017dynamic,thomas2022mobilise,mccarthy1977resource,mcadam1999political,klandermans1984mobilization}.

Yet, critical puzzles remain. First, the conditions under which instances of protest build into an enduring, sustained movement differ from those which generate one-off protests~\cite{selvanathan2020marches,paret2021persistent,saunders2012explaining,tarrow1993cycles,tilly2015social}. Participation in collective action is a goal-directed behaviour but social change takes time, implicating the repeated efforts of people over months, years, and decades \cite{de2013dramatic,tarrow1993cycles,mcadam1999political,klandermans1987potentials}. Other evidence suggests that the factors that shape incidental action may be distinct from those factors that shape when people \textit{keep acting} \cite{cohen2023should}  and \textit{how} they continue to engage \cite{louis2020volatility,meyer2004conceptualizing,tarrow1993cycles}. This paper addresses this critical gap, asking: what are the conditions that foster an enduring, sustained social movement?  

Tackling the explicitly temporal dimensions to such movements helps to address another puzzle. Social movements typically display distinct balances of conventional and radical tactics~\cite{roggeband2017handbook,moskalenko2009measuring,mccauley2017understanding,della2008research}. \textit{Conventional actions} (e.g., petitioning, peaceful demonstrations, lobbying) are those that are common and approved of while \textit{radical actions} (e.g., illegal actions, sabotage, civil disobedience) are those that are unusual and undesirable from the perspective of authorities \cite{louis2020volatility, wright1990responding}. Accordingly, a second goal of our analysis is to identify the conditions under which radicalism may emerge as a preferred tactic as part of a longer-term struggle to achieve the goals of the movement (see also \cite{becker2017dynamic,tausch2011radical,mccauley2017understanding,tilly2015social,tarrow1993cycles}). We ask: What are the conditions under which people will come to adopt more radical forms of action in pursuit of their movement’s goals? 

Addressing these questions about the conditions for enduring action and the emergence of radicalism requires a theoretical analysis of the effects of repeated interactions between individual protesters, social movements and the authorities to whom they appeal for change (i.e., cycles of protest \cite{benford2000framing, snow1992master, tarrow1993cycles,tilly2015social,saunders2012explaining,paret2021persistent}). We draw on the insights of a current psychological model of collective action outcomes, the Disidentification, Innovation, Moralisation, Energisation (DIME) model \cite{louis2020volatility, louis2022failure, lizzio-wilson2021collective}. As we outline below, the DIME Model explains dis/engagement and radicalism as a function of the interactions between members of a social movement and the responsiveness versus intransigence of authorities to whom protesters appeal for change.  

Methodologically, we require an approach that can capture the effects of time and scale (i.e., reflecting the repeated interactions of a whole movement). Agent-based models (ABM) represent one solution \cite{Miller2009}. ABMs are perhaps especially useful when it is implausible to test the effects of repeated interactions over extended periods of time, involve population-level phenomena, difficult-to-access populations (e.g., activists, radicals), and/or ethically fraught contexts (e.g., repressive contexts where it may not be safe to use alternative methods). Accordingly, we implemented and analysed the Disidentification Innovation Moralisation Energisation Model of Collective Action Outcomes Simulation (DIMESim). DIMESim is a theoretically-founded \cite{louis2020volatility} and empirically-informed \cite{louis2022failure} agent-based model of collective action. Our findings show that where an authority is responsive, and people individually and collectively contest failure, a moderate-sized conventionally engaged movement persists. Alternatively, where an authority repeatedly fails the group, and the protesters are unable to individually or collectively re-frame failure as success, then the movement population becomes inactive but radicalised (latent radicalism).

\section*{Modelling Cycles of Collective Action}

Our approach draws, inter alia, from the interdisciplinary literature on the drivers of participation but differs from the established tradition of psychological research into collective action \cite{agostini2021toward,thomas2022mobilise,becker2017dynamic} in focusing on a population who have already engaged in action. One of the best predictors of future behaviour is past behaviour \cite{ouellette1998habit} and, indeed, the literature confirms that prior participation predicts future action \cite{thomas2020testing}. However, there are clearly instances in which people will temporarily or permanently disengage~\cite{saunders2012explaining,nepstad2004persistent}, and other instances where people will change tactics from conventional to radical actions (and vice versa; see \cite{moskalenko2009measuring, mccauley2017understanding, becker2017dynamic}). The Disidentification, Innovation, Moralisation and Energisation model of collective action outcomes (DIME Model; \cite{louis2020volatility}) presents a theoretical framework for how interactions between authorities and a movement shape ongoing dis/engagement, as well as tactics (conventional actions versus radical actions), over time. The DIME Model posits that when a movement is experiencing success in achieving its goals, that success will reinforce tactical choices – notably, people that have previously adopted conventional tactics will continue to do so.


On the other hand, the DIME Model posits that when a movement is struggling to achieve its aims, diverging trajectories will emerge. One possible outcome is that, in the face of intransigence and failure, people will disidentify (give up, walk away \cite{becker2014group,mccarthy1977resource,meyer2004conceptualizing}). Thus, when a movement repeatedly experiences failure, it is expected that some adherents will disengage and become inactive. It is also possible that actors will decide that, in the face of repeated failure, new forms of action are necessary (Innovation). For those who had taken prior conventional action, this would mean adopting new radical methods (see \cite{moskalenko2009measuring,della2018radicalization,meyer2004conceptualizing,mccarthy1977resource}). Conversely, for those who had previously adopted radicalism, innovation represents a shift back to more conventional (accepted) forms of action (see \cite{jones2008terrorist,della2018radicalization,meyer2004conceptualizing,mccarthy1977resource}). Finally, another possibility is that supporters will see the struggle as of increased moral worth (\cite{skitka2021psychology,nepstad2004persistent}; i.e., Moralisation) and pursue their previous forms of action with renewed vigour and energy (Energisation). In this way, the success and failure of prior tactics (i.e., conventional action versus more radical forms of action) shapes subsequent tactics. People may give up on immediate action (disengage), double down on conventional actions or increase support for more radical action \cite{louis2020volatility, tausch2011radical}. 

Indeed, experimental tests of the DIME propositions showed that each of these outcomes occurs \cite{louis2022failure}, though for different sub-groups of people \cite{lizzio-wilson2021collective}. Louis and colleagues \cite{louis2022failure} report a meta-analysis of nine experiments ($N =1663$) across different movements (environment, pro/anti-immigration, pro/anti-abortion, marriage equality). Participants were exposed to the ostensible success or failure of their movement, in a context where they were asked to imagine that they had engaged in either conventional or radical tactics. The tests revealed that people who are experiencing failure are more likely to disidentify, especially if they have been acting radically; that people who are experiencing the failure of their conventional tactics are more likely to innovate; and that people who are experiencing failure are more likely to endorse radical tactics, as are those who have been more radical in the past. What is not known is how these effects iterate over time, where protesters may experience a win one day, and a setback the next. An as-yet untested insight of DIME is that failure will produce diverging trajectories in the movement population overall. 

It is also the case that success and failure must be interpreted, and people differ in their perceptions of what constitutes failure and success \cite{hornsey2006why,snow1992master,meyer2004conceptualizing,mccarthy1977resource,nepstad2004persistent,bursztyn2021persistent}. For some, failure may be contested and re-framed as success~\cite{cohen2023should,snow1992master,nepstad2004persistent}: ``our action did not effect policy change but we built an oppositional movement". This cognitive, \textit{individual re-framing} tends to occur among people who are more strongly committed to the cause, i.e., people high in identification or psychological commitment to the group~\cite{hornsey2006why,blackwood2012if}. Members of social movements are also influenced by fellow supporters' views about the progress of the movement \cite{thomas2014social,snow1992master}, allowing for group members to socially, \textit{collectively re-frame} failure in ways that promote continued engagement with the movement\cite{gulliver2021psychology,drury2005explaining}: ``we agree that although we did not effect change, we stood up for our values''.  

\section*{From Theory to Simulation: Introducing DIMESim}

As noted above, ABMs are well suited to modelling emergent properties of complex systems involving interactions between individuals (protesters), external factors (i.e., authorities who signal success and failure) and social networks (fellow protesters), over time~\cite{Miller2009}. Since the seminal work of Granovetter~\cite{granovetter1978threshold}, there have been many ABMs of collective action and protests put forward in political science~\cite{siegel2011does,lohmann1993signaling,kuran1989sparks}, economics~\cite{acemoglu2008persistence,rubin2014centralized}, sociology~\cite{myers2000diffusion,centola2013homophily,macy2020threshold}, and more~\cite{epstein2002modeling,moro2016understanding,ye2021collective}. A significant body of literature has focused on economical, utilitarian or institutional factors for collective action and regime change~\cite{lohmann1993signaling,acemoglu2008persistence,makowsky2013agent,rubin2014centralized,centola2013homophily}. A game-theoretic framing is often used, with agents aiming to maximise individual or collective utility. Other models, inspired by epidemic or complex contagion processes, focus on the role social influence and coordination in creating cascades of protest movements~\cite{ye2021collective,epstein2002modeling,moro2016understanding,siegel2011does,kuran1989sparks,granovetter1978threshold,myers2000diffusion,macy2020threshold}. Additional factors of interest can be incorporated, such as the presence of an authority~\cite{makowsky2013agent,acemoglu2008persistence,lohmann1993signaling}, suppression of protesters~\cite{epstein2002modeling,siegel2011does,moro2016understanding}, and the impact of personal preferences~\cite{kuran1989sparks,siegel2011does,rubin2014centralized}. 

DIMESim shares some similarities with existing models (i.e. the presence of a network, authority, and threshold-type dynamics for behaviour) but departs from these prior approaches. Critically, DIMESim is both theoretically-founded on the DIME model, and empirically-informed by associated experimental data -- that is, the model is parametrised based on evidence from human experiments~\cite{louis2022failure}. DIMESim focuses on modelling how an agent's psychological states evolve through interaction with the authority and fellow protesters, and how these states determine the tactics adopted~\cite{louis2020volatility,louis2022failure}. It therefore goes beyond the binary distinctions of dis/engagement that characterise the current literature, to include (conventional, radical) tactics, reflecting the diversity of real-world movements.

Bringing together these methodological and theoretical advancements, we developed a theoretically informed, empirically evidenced agent-based model (DIMESim) to simulate effects of interactions between protesters, authorities, and supporters over time and at scale. As we show below, incorporating these features into a computational model allows for the emergence of complex collective phenomena in ways that would elude traditional research methodologies (interviews, surveys, experiments) but provide powerful tests of propositions about the conditions for an enduring movement and emergent radicalism~\cite{jackson2017agent}. DIMESim itself provides a unified modelling framework for simulating the diversity of real-world movements, from public disengagement to widespread action, conventional and radical movements.


\section*{The DIMESim Model}

\begin{figure}[h!]
	\centering \includegraphics[width=0.8\textwidth]{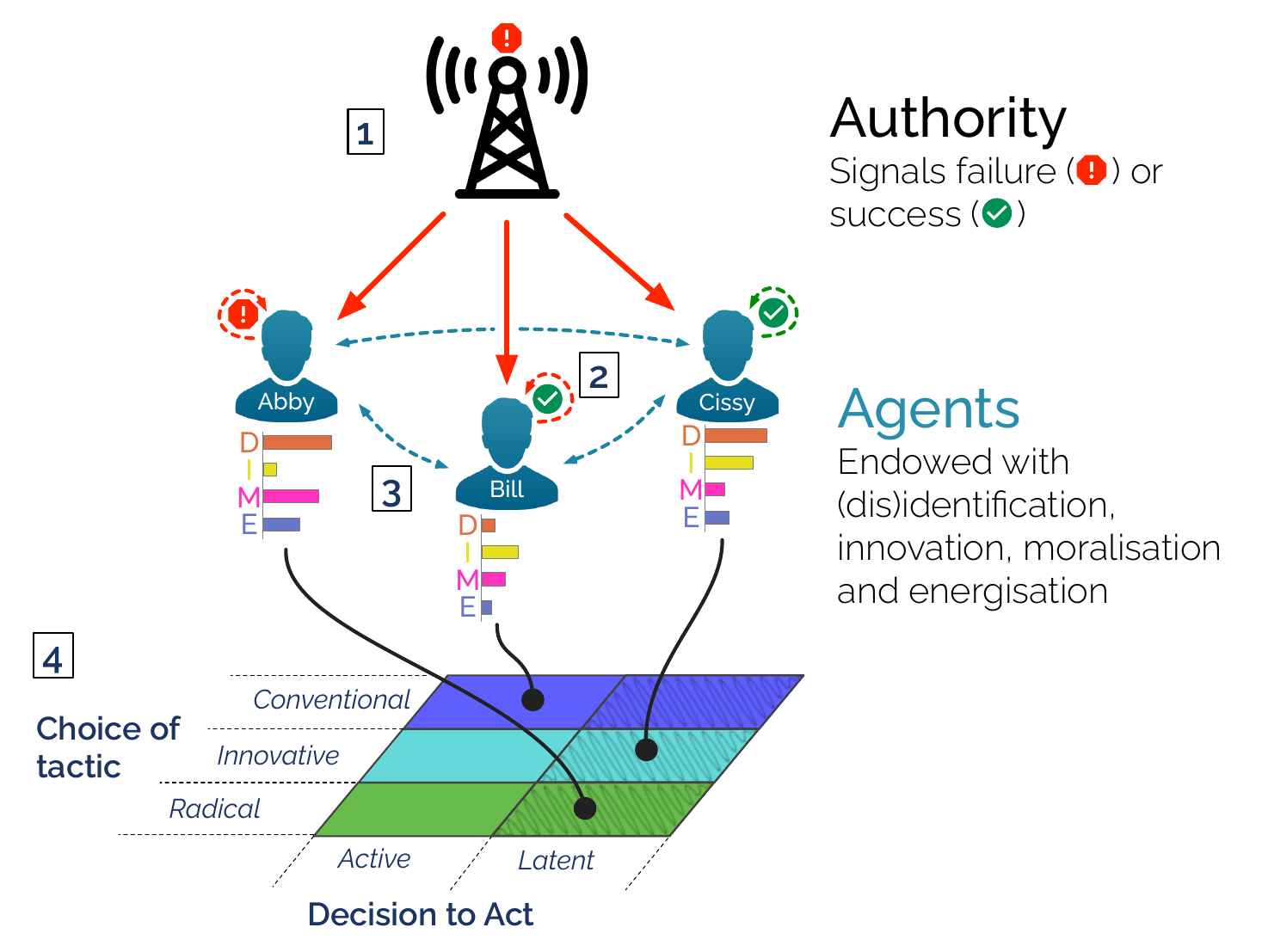}
	\caption{{\bf Schematic overview of four phases of the DIMESim model.} First, an Authority broadcasts to all agents a common success or failure signal relating to their demands (1). The signal is received by each agent who can (2) individually re-frame the signal from failure to success (e.g. shown by Bill and Cissy), and, in a third step, (3) collectively re-frame based on the success/failure interpretations of neighbours in the social network. Finally (4), based on the perceived signal and the agent's DIME variable states, the agent will make a decision about whether to act (active or latent), in either a conventional, innovating, or radical direction. In the schematic example, all agents have received a failure signal (red arrow). Abby interprets the failure signal as a failure and does not individually or collectively re-frame that failure as success (red exclamation). Bill collectively re-frames failure as success (green tick). Cissy individually re-frames failure to perceive the outcome as a success (green tick). } \label{fig:ModelSchematic}
\end{figure}

The DIMESim agent-based model is 
represented in Fig.~\ref{fig:ModelSchematic}. The complete mathematical details are presented in Methods and Materials. Agent-based models have three components: the agents, the environment, and the interactions, which we describe in turn. DIMESim contains a population of $n\geq 2$ agents each representing an individual protester, indexed as agent $i = 1,2, \hdots, n$. The agents comprise the mobilisation potential of a movement~\cite{klandermans1984mobilization,mccarthy1977resource}, where motivational and structural factors influence who acts, how they do so, and when. The dynamics play out over a sequence of discrete time steps $t=0,1,2,\hdots$. 

Each agent is endowed with social psychological attributes of level of commitment to the cause (Dis / identification) and moral conviction, as well as propensity to innovate and energise, based on the sample estimates in the empirical tests \cite{louis2022failure}. Specifically, the states for an arbitrary agent~$i$ are $D_i(t)$, $I_i(t)$, $M_i(t)$, $E_i(t)$ and $x_i(t)$, representing an agent's Disidentification, Innovation, Moralisation, Energisation, and protest action, respectively. 

The primary feature of the environment is an authority, whom the protesters appeal to, and who responds to each social change attempt by sending out a signal that the protest event has succeeded or failed. Interactions between the protesters and authorities, and within the movement, are modelled in a sequence of three processes. 

The first process involves the authority sending out a common signal to each agent: at every time step, the probability of a failure signal and success signal is $p\in [0,1]$ and $1-p$, respectively -- that is, higher $p$ represents more failure signals. This is akin to the authority using state-controlled or mass media media to broadcast the outcome. In highly repressive regimes, there is little scope for further contestation~\cite{pierskalla2010protest,moghaddam2020psychology}. Yet, we know that what is seen to be a success and a failure are subjective and people can contest whether a given outcome is a success or failure both individually~\cite{hornsey2006why,lizzio-wilson2021collective} and collectively through social interaction with fellow protesters (see~\cite{thomas2014social,drury2005explaining,blackwood2012if,snow1992master,meyer2004conceptualizing,della2018radicalization}). Incorporating these important aspects allows for a more ecologically valid test of these outcomes than does assuming that people and movements passively accept the signals of authorities. These two observations are captured by the second and third processes, respectively.

In the second process, individual contestation occurs, whereby each agent has a probability to contest a failure signal from the authority and re-frame it as a success. Following \cite{hornsey2006why, drury2005phenomenology, gulliver2021psychology} this is especially likely to be the case for highly identified members. In DIMESim, the probability of re-framing increases as the level of disidentification decreases (as $D_i(t)$ decreases) and as a threshold parameter $F\in[0,1]$ decreases. 
Thus, decreasing $F$ captures a protester that more easily re-frames because it reflects a lower threshold for re-framing.   

The third process captures collective re-framing. After agents individually contest the failure signal, they then share their perception of whether or not the protest was successful with other agents. In DIMESim, the sharing occurs in a fixed network structure, analogous to a social media network. If agent~$i$ believes the protest has failed at the start of the collective re-framing process, they can become convinced of its success when a threshold fraction $\phi\in [0,1]$ of their network neighbours believe the protest is successful. An integer parameter $R \geq 1$ captures how long this collective process goes on for; the larger is $R$, the more agents are able to freely communicate success and failure signals with each other. The re-framing processes do not occur when the authority sends out a success signal; i.e. all agents will perceive the signal as success.

The next step of the model involves each agent internalising the perceived outcome of the protest – that is, the perceived signal resulting from the sequence of the authority signal, the individual re-framing and the collective re-framing. First, agent~$i$ updates its DIME variables, 
taking into account i) agent~$i$'s belief in the failure or success of the protest movement and ii) the agent's current action $x_i(t)$, with the coefficients of the equations derived from the empirical data in~\cite{louis2022failure}. 

Finally, in the fourth step, an agent will decide on its next action, $x_i(t+1)$. This comprises two aspects: their decision to act and choice of tactics. An agent's engagement depends on whether they are sufficiently identified with the cause~\cite{agostini2021toward}. Namely, agent~$i$ will chose \textit{not to protest} if $D_i(t+1)$ is greater than the average of $I_i(t+1)$, $M_i(t+1)$ and $E_i(t+1)$.
Otherwise, agent~$i$ will protest at time $t+1$. An agent's tactical choice is either conventional or radical, and depends on their last active protest tactic, and whether they wish to innovate at the current time step.


This sequence of events (see Fig.~\ref{fig:ModelSchematic}), from authority signal to agent updating of their DIME variables and action, occurs in a single time step. The computational model then plays out over a sequence of time steps $t=0,1,2,\hdots$. The primary parameters of the model are summarised in Table~\ref{tab:parameters_primary}.

\begin{table}[h!]
    \centering
    \caption{The four primary parameters in DIMESim. The results interrogate outcomes of the  model over the space of these parameters.}
    \begin{tabular}{c | c | c | p{0.45\textwidth}} 
         Parameter & Range & Short-hand identifier & Interpretation \\ \midrule
         $p$ & $[0, 1]$ & Probability of failure signal from authorities& Low values indicate that the movement is succeeding, while high values indicate that the movement is failing.\\
         $F$ & $[0 , 1]$ & Threshold for individual cognitive re-framing & As $F$ increases, it becomes more difficult to individually re-frame failure as a success.\\
         $\phi$ & $[0, 1]$ & Threshold for collective social re-framing & As $\phi$ increases, it becomes more difficult to collectively re-frame failure as a success.  An agent requires a greater fraction of its network neighbours to be perceiving success, in order for the agent to also perceive success.\\
         $R$ & $\geq 1$ & Length of free communication & As $R$ increases, protesters are able to more freely communicate with each other during the collective re-framing stage.
    \end{tabular}    
    \label{tab:parameters_primary}
\end{table}


An important feature of DIMESim is that an agent can choose not to protest but still have an orientation in terms of which tactic the agent would adopt if they were to protest. Recent models distinguish between latent (e.g., attitudinal) and active (e.g., behavioural) support for a movement~\cite{mccauley2017understanding}. Populations can appear inactive (disengaged from the public sphere)~\cite{snow1992master,klandermans1984mobilization} but will re-engage when the social and psychological conditions are right, because actors did not attitudinally disengage. It is also the case here that agents can exist in a state of \textit{innovation}, alternating their choices of actions between radical and conventional (consistent with evidence presented in~\cite{thomas2019vegetarian,pavlovic2024classes, jones2008terrorist}. Agent~$i$ will innovate if agent~$i$'s innovation exceeds the average of their moralisation and energisation.
If agent~$i$ innovates, they will change their action relative to the action they took when last active, i.e. from conventional to radical, or from radical to conventional. 

Bringing together combinations about the decision to act/not (latent or active); choice of tactics (conventional, radical, innovative) yields a typology of different types of protesters (see Table~\ref{tab:action_types}). Agents are denoted as \textit{latent radical} or \textit{latent conventional} if they are inactive, are not innovating, and are orientated towards radical or conventional tactics, respectively. Agents are \textit{latent innovators} if they are inactive and innovating (and thus alternating between radical and conventional tactics). Similarly, an agent is an \textit{active radical},  \textit{active conventional} or \textit{active innovator} if they are active, and adopting radical, conventional or innovative tactics, respectively.

\begin{table}[h!]
    \centering
    \caption{Types of actions that an agent in the DIMESim model can display.}
    \begin{tabular}{c | c c } 
         Agent Type & Intention to Act & Choice of Tactic \\ \midrule
         Active conventional agent & Active & Conventional \\
         Active innovator & Active & Innovating \\
         Active radical agent & Active & Radical \\         
         Latent conventional agent & Inactive & Conventional \\
         Latent innovator & Inactive & Innovating \\
         Latent radical agent & Inactive & Radical 
    \end{tabular}    
    \label{tab:action_types}
\end{table}

\section*{Results}
Our simulations involve modelling the effects of interactions with authorities ($p$; probability of failure signals), at increasing levels of individual ($F$; threshold parameter) and collective re-framing ($\phi$; threshold parameter), and varying lengths of free communication ($R$); see Table \ref{tab:parameters_primary}). 
Given the complexity in the design, we first present results under two different idealised scenarios that reflect combinations of authority signals, individual and collective re-framing, over time. We then examine more systematically the effects of (degrees of) success/failure and individual re-framing whilst holding collective re-framing steady. Finally, we probe the effects of collective re-framing at different levels of $\phi$ and $R$.

Full details of the simulation parameters are provided in the Methods and Materials. Here, we briefly summarise the key aspects that are uniform across the simulations. We initialise all agents to be active and protesting conventionally, $x_i(0)=1$ for all $i=1,2,\hdots, n$, while $D_i(0)$, $I_i(0)$, $M_i(0)$, and $E_i(0)$ are drawn from distributions estimated from empirical data. Due to the stochastic nature of the model, we conducted $20$ Monte Carlo simulations out to $T = 10,000$ time steps and compute average outcomes across the $20$ replicates. The agents interact over a synthetic network constructed using the Holme-Kim algorithm~\cite{holme2002network}, which generates networks with the properties of real-world social networks. Unless noted otherwise, we fixed $R = 10$, allowing agents to communicate success or failure signals with each other for a substantial period during the collective re-framing phase, prior to the next authority signal.

\subsection*{Emergent outcomes of two idealised scenarios} 

\begin{figure}[h!]
 \centering
 \includegraphics[width=\textwidth]{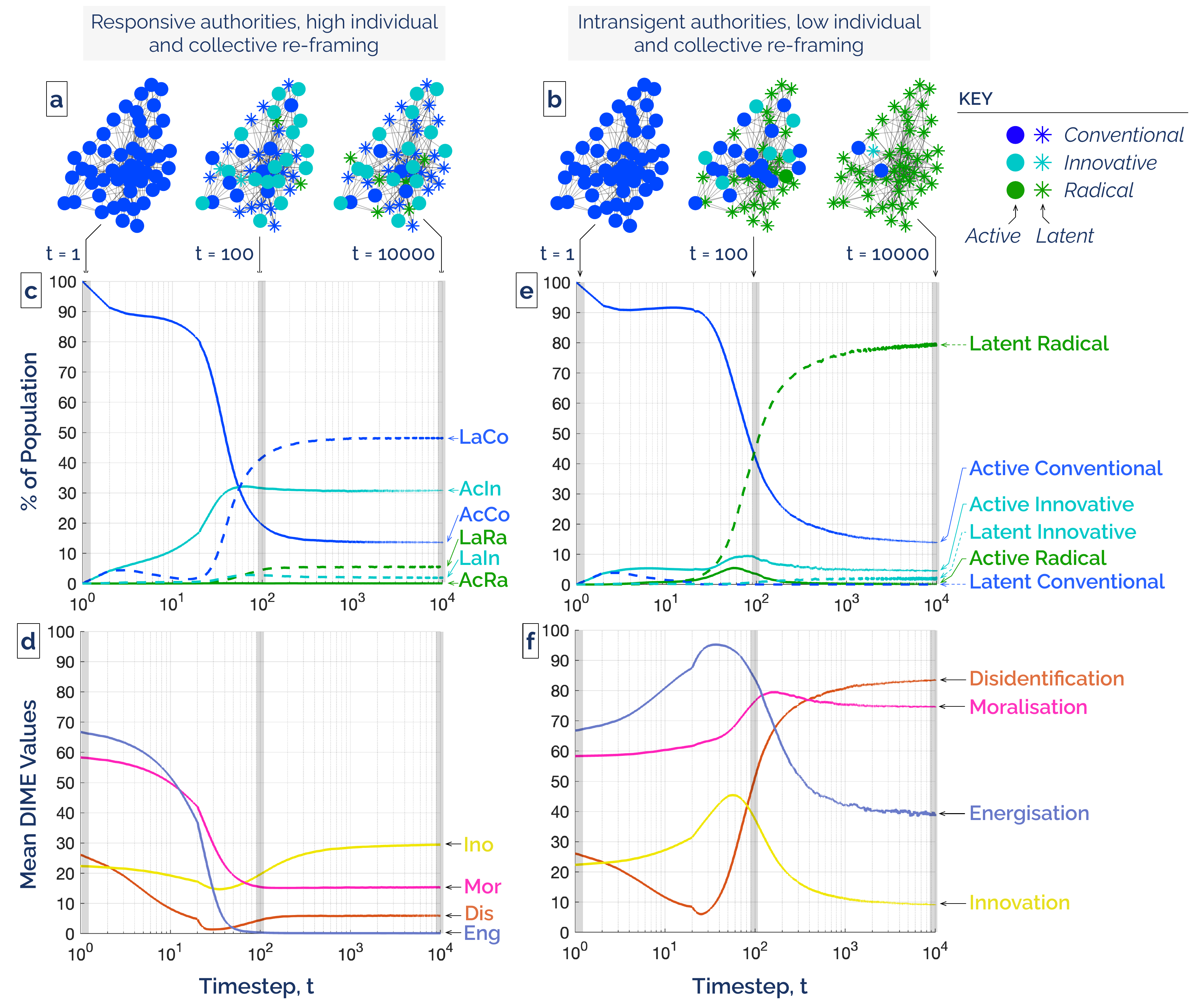}
 \caption{{\bf The emergence of very different outcomes under conditions of responsive vs. intransigent authorities, individual and collective re-framing} (a),(b): Network examples ($n=50$) at $t\in\{1,10^2,10^4\}$ and, (c)-(f): aggregate timeseries plots ($n=1000$, $t=1\dots 10^4)$ of population composition for different protester types and their mean DIME variables for two parameter regimes. Left panels (a, c, d): High movement success, high individual + collective re-framing, $p=0.2, F=0.2, \phi = 0.2$. Right panels (b, e, f): High movement failure, low individual + collective re-framing, $p=0.8, F=0.8, \phi = 0.8$. In (c), we see a diverse range of protester types, with latent conventional (LaCo), active innovators (AcIn), and active conventional (AcCo) being most prevalent. In (e), the population is comprised almost entirely of latent radical protesters, i.e. they are inactive but orientated to radical tactics if they were to become active.}
 \label{fig:NetworkTimeSeriesComposite}
\end{figure}

Figure~\ref{fig:NetworkTimeSeriesComposite} displays the outcomes under two idealised scenarios. The first, shown in the left panels of the figure (Fig.~\ref{fig:NetworkTimeSeriesComposite}(a), (c) and (d)), is where the authority is responsive $p = 0.2$ (high probability of success signals), agents have a high chance of individually re-framing a failure signal to a success signal $F = 0.2$, and an agent is easily influenced by the success signals of their neighbours in the collective re-framing stage $\phi = 0.2$. The second, shown in the right panels of the figure, (Fig~\ref{fig:NetworkTimeSeriesComposite}(b), (e) and (f)) corresponds to a scenario of an intransigent authority (high probability of failure signals) with $p = 0.8$, agents have a low chance of re-framing failure as success, $F = 0.8$, and are not easily influenced by the success signals of group members in the collective re-framing stage, $\phi = 0.8$. In panels (a) and (b), we show a small $n = 50$ network to highlight the topological features of the model under each scenario, while time-series obtained in panels (c), (d), (e) and (f) shows the aggregated simulation results for a larger $n=1000$ network; both display the evolution of effects over time (i.e., between timestep $1-10,000$).

Fig.~\ref{fig:NetworkTimeSeriesComposite} reveals two very different emergent phenomena. The outcomes in the responsive authority scenario (Fig.~\ref{fig:NetworkTimeSeriesComposite}: panels (a), (c) and (d)) reveal a trajectory of sustained conventional engagement. Panel (c) suggests that as $t\to T$ (i.e. at steady state), three types of protesters are most prevalent in the population: $48\%$ latent conventional  (sympathisers in Klandermans' terms \cite{klandermans1987potentials}), $31\%$ active innovators, and $14\%$ active conventional. In other words, virtually all inactive agents favour conventional tactics, while those that are actively protesting are either innovators, or conventional protesters. Panel (d) shows the average values of the DIME variables in the population. On average, agents are highly identified with the protest movement (low disidentification), have low-moderate levels of innovation, and low moralisation and low energy. This scenario resembles the sustained and emergent collective action that is often predicted in existing agent-based models~\cite{granovetter1978threshold,epstein2002modeling,siegel2011does}, but our model is unique in explaining these outcomes as a function of the psychological state of agents and their interactions with the authority. 


In contrast, the intransigent authority scenario (Fig.~\ref{fig:NetworkTimeSeriesComposite}: panels (b), (e) and (f)) shows very different emergent phenomena under conditions of high failure, and low individual and collective re-framing. Here, the simulation revealed the emergence of mass latent radicalism, whereby the overwhelming majority of the population (79\%) have become inactive radicals – agents who are inactive but orientated towards radical protest actions if triggered (see Table 1). A small proportion (14\%) of agents continue to act conventionally. Fig.~\ref{fig:NetworkTimeSeriesComposite}(f) shows that agents are on average highly disidentified with the protest movement, highly moralised, have moderate energisation, and low innovation.

\subsection*{Interactions between the authority and individual re-framing}


\begin{figure}[th!]
	\centering 
    \includegraphics[width = 0.9\textwidth]{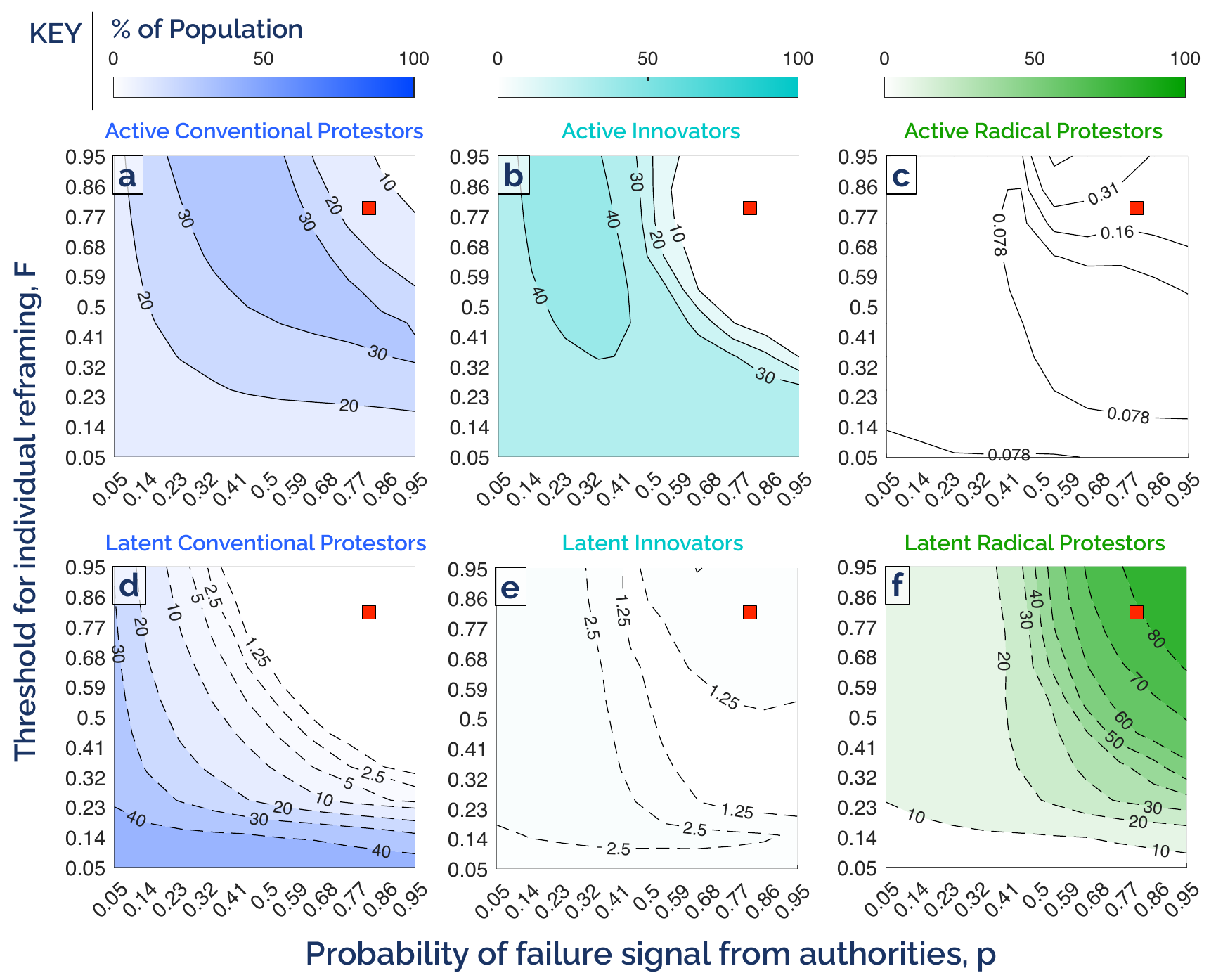}
    \caption{{\bf Mapping the emergence of the six protester types under varying probability of failure signals ($p$) and threshold for individual re-framing ($F$).} Contour plots show population composition (numbered lines, colourbars) at the long-run steady state for the six different agent types under study, aggregating simulations at a given ($p,F$) location ($n=1000$): active conventional (a), active innovators (b), active radical (c), latent conventional (d), latent innovators (e) and latent radicals (f). Collective re-framing is held constant across plots ($\phi = 0.8, R = 10$). The red square marker indicates the intransigent authority context studied in the right panels of Fig.~\ref{fig:NetworkTimeSeriesComposite} ($p=0.8, F=0.8, \phi = 0.8, R = 10$). Contours show an interaction between probability of failure signals ($p$) and threshold for individual re-framing ($F$).}
    \label{fig:ParameterSweep_pF}
\end{figure}

\begin{figure}[th!]
	\centering 
    \includegraphics[width = 0.8\textwidth]{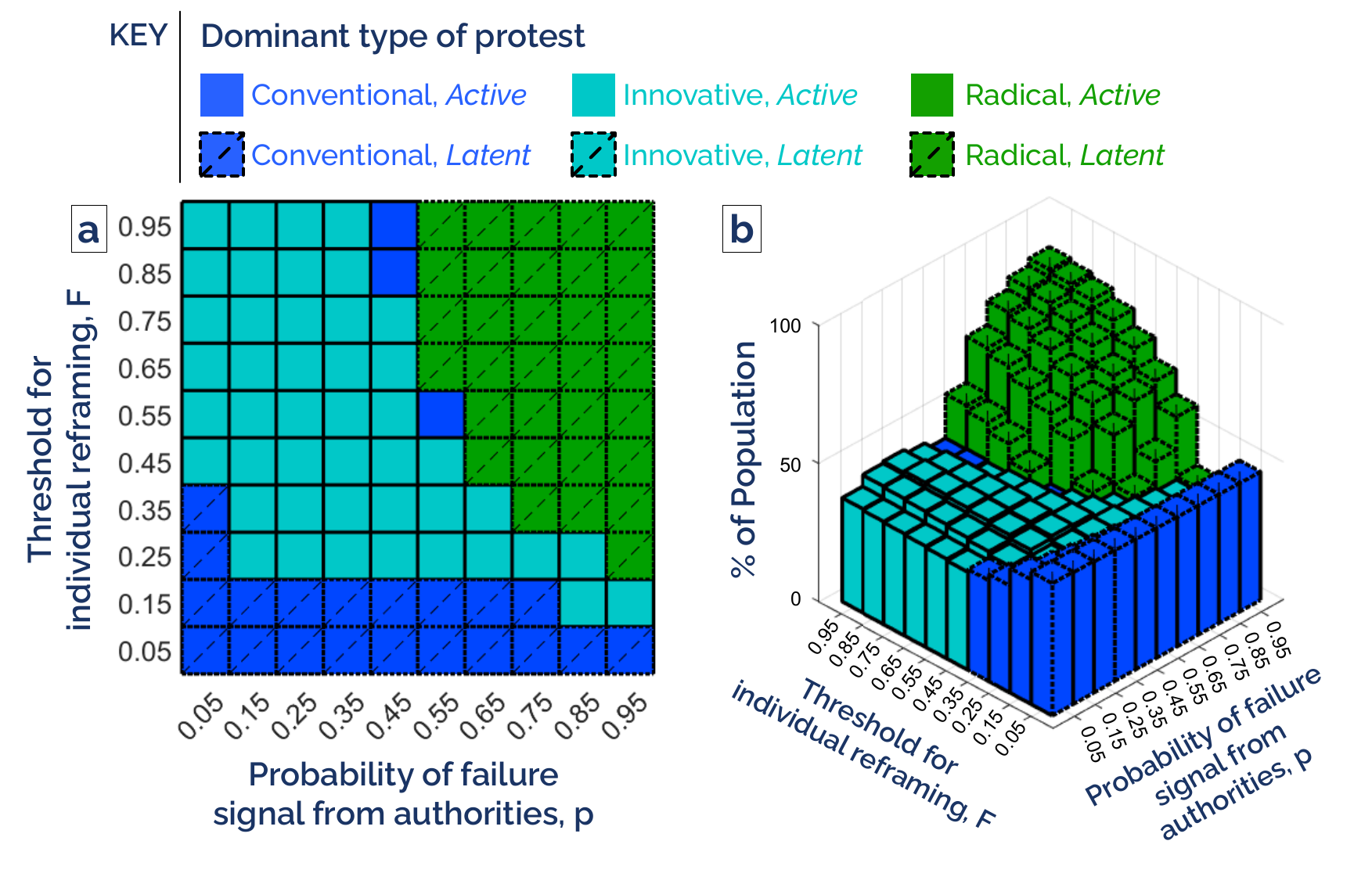}
    \caption{{\bf Dominant types of agent under varying probability of failure signals ($p$) and threshold for individual re-framing ($F$).} (a): Variation in dominant action, utilising colour and hatching (see Key) for the six agent types, with $n=1000$ and varying probability of failure signals ($p$) and threshold for individual re-framing ($F$) over [0,1] (holding collective re-framing at $\phi = 0.8$ and $R = 10$). (b): Population composition (on z-axis) for the corresponding dominant strategy at each parameter pair. The latent radical dominant region shows the highest population fraction of dominant action of any regime.}
    \label{fig:DominantActions_pF}
\end{figure}

We next systematically interrogated the  probability of failure signal $p \in [0, 1]$ and threshold for individual re-framing $F \in [0, 1]$, on the distribution of agent types (see Table~\ref{tab:action_types}), holding collective re-framing constant at $\phi = 0.8$ and $R = 10$. Fig.~\ref{fig:ParameterSweep_pF} presents the average population fraction for each agent at steady state. The different colours in Fig.~\ref{fig:ParameterSweep_pF} correspond to choice of tactics (conventional, innovating, or radical); the top panel (Fig.~\ref{fig:ParameterSweep_pF}(a)-(c)) denotes \emph{active} strategies whilst the bottom panel (Fig.~\ref{fig:ParameterSweep_pF}(d)-(f)) denotes \emph{latent} strategies. Plots of DIME variables of the population for each protester type are available in the Supplementary Materials. Fig.~\ref{fig:DominantActions_pF} complements Fig.~\ref{fig:ParameterSweep_pF} to present an overview of the dominant type of protester at each combination of $p$ and $F$ (Fig~\ref{fig:DominantActions_pF}(a)) and also as a function of the population fraction (Fig.~\ref{fig:DominantActions_pF}(b)).

Fig.~\ref{fig:ParameterSweep_pF} shows that active radical agents (Fig.~\ref{fig:ParameterSweep_pF}(c)) and latent innovators (Fig.~\ref{fig:ParameterSweep_pF}(e)), occupy a consistently marginal presence at all levels of the simulation, never exceeding 4\% of the population. On the other hand, when failure $p$ and $F$ are both high (e.g. $p \geq 0.55$ and $F \geq 0.55$, denoting low re-framing), latent radical agents comprise over 60\% of the population, and up to 87\% (Fig.~\ref{fig:ParameterSweep_pF}(f), Fig.~\ref{fig:DominantActions_pF}(a)) – the most prevalent majority response for any of the simulations (denoted by the green bars in Fig.~\ref{fig:DominantActions_pF}(b)). The results in Fig.~\ref{fig:ParameterSweep_pF}(f) suggest that latent radicalism does not emerge due to a single factor, but is an outcome of multiple factors. Even when failure signals are high, if there is capacity to individually (cognitively) re-frame and perceive that failure as success (i.e., $F$ is low), then latent radicalism will not emerge. Similarly, when the protesters lack capacity to re-frame (high $F$), but the authority is responsive (low $p$) then latent radicalism will emerge amongst only a small proportion of the population.

Indeed, when $p$ and $F$ are medium-to-low in value, the population is far more heterogeneous (Fig.~\ref{fig:DominantActions_pF}(b), consistent with \cite{lizzio-wilson2021collective}) and three types of agents are typically observed: active conventional (Fig.~\ref{fig:ParameterSweep_pF}(a)), active innovators (Fig.~\ref{fig:ParameterSweep_pF}(b)), and latent conventional (Fig.~\ref{fig:ParameterSweep_pF}(d)). When $p$ and $F$ are both low ($\leq 0.15$), latent conventional agents are the majority (Fig.~\ref{fig:DominantActions_pF}), forming 44\% on average and up to 48\% of the population. These results suggest that a population of sympathetic but largely inactive actors emerges from situations where the movement is actually and understood to be (re-framed as) succeeding (Fig.~\ref{fig:ParameterSweep_pF}(d)). 
On the other hand, active innovation – people oscillating between conventional and more radical tactics from action to action (as observed by~\cite{thomas2019vegetarian}) – was a prominent strategy (39\% on average, up to 43\% of the population; Fig.~\ref{fig:ParameterSweep_pF}(b)), except for at high levels of failure and low re-framing (see Fig.~\ref{fig:DominantActions_pF}). Indeed, even a relatively low proportion of failure signals from an authority ($F \geq 0.15$) will be associated with innovation, if there are constraints on the ability of protesters to individually re-frame failure as success (i.e., as $F$ increases; Fig.~\ref{fig:ParameterSweep_pF}(b)). 
By contrast, active conventional protesters were prevalent across nearly all levels of the simulations (37\% on average) but never exceeded 39\% of the total population (Fig.~\ref{fig:ParameterSweep_pF}(a)). Active conventional protesters were not the majority or dominant response within the population (Fig.~\ref{fig:DominantActions_pF}) but were most prevalent at moderate levels of failure and re-framing (Fig.~\ref{fig:ParameterSweep_pF}(a)).

\subsection*{The role of collective contestation: re-framing and interaction} 

\begin{figure}[th!]
    \centering
    \includegraphics[width=0.9\textwidth]{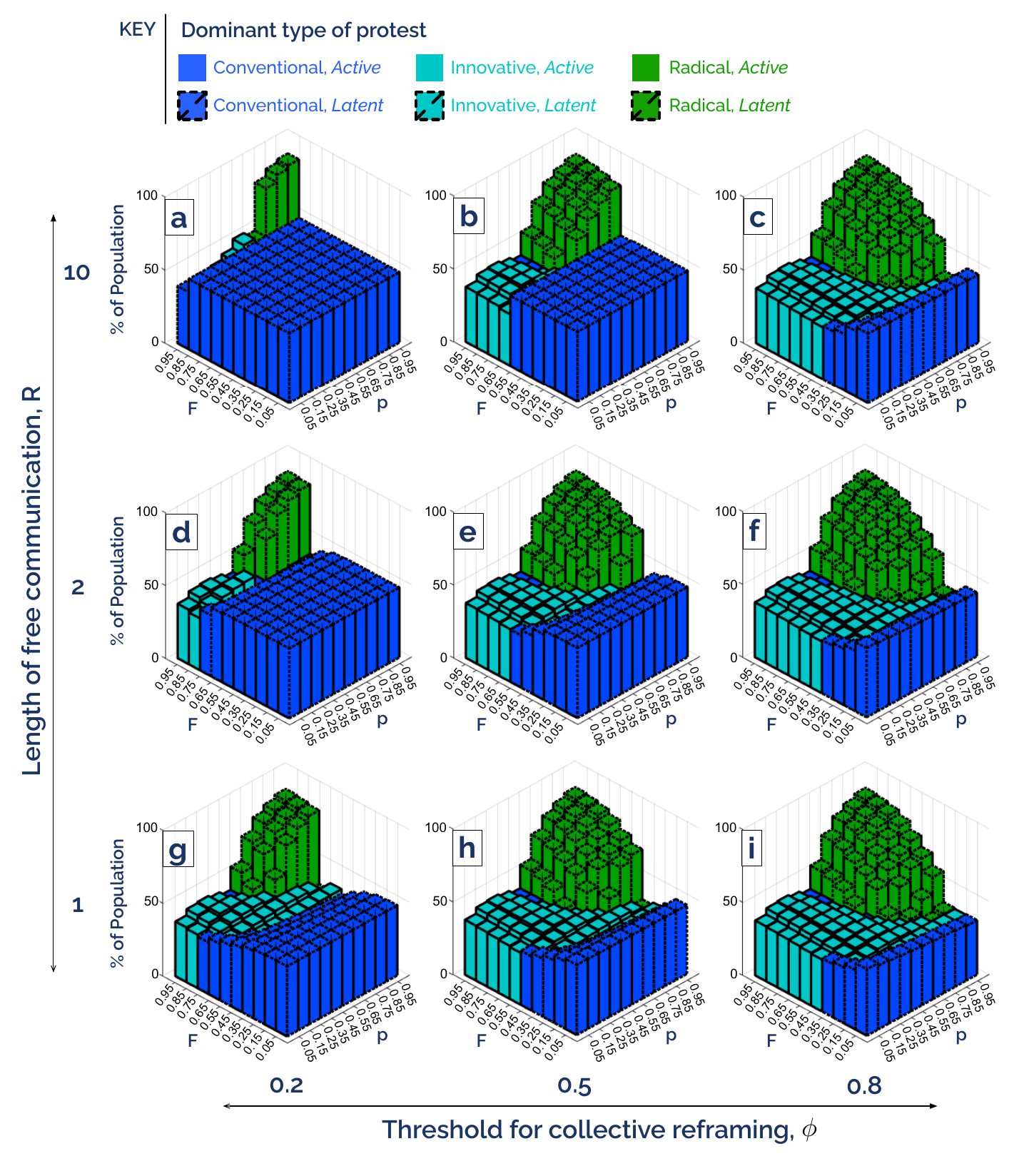}
    \caption{{\bf The role of collective re-framing in the emergence of dominant protest types.} Each panel depicts the population composition of the dominant action in the steady state, for a given probability of failure signal---threshold for individual re-framing pair ($p,F$) ($n=1000$). Rows and columns show the outcome when varying: (left to right) the threshold for collective re-framing ($\phi \in \{0.2, 0.5, 0.8\}$); and (bottom to top) the length of free communication ($R \in \{1, 2, 10\}$). Colouring follows the key given, covering the six agent types under study. Overall, we can see the largely resilient nature of the dominant latent radical region, under variation in collective re-framing.}
    \label{fig:DomActions_Rphi}
\end{figure}



Finally, we explored the influence of the collective re-framing process which is shaped by two factors: threshold parameter $\phi \in [0, 1]$ and length of free communication $R \geq 1$. Fig.~\ref{fig:DomActions_Rphi} shows the population fraction of the dominant action for each probability of failure signal---threshold for individual re-framing pair ($p,F$). From the left to right columns, $\phi$ decreases; i.e. the threshold for collective re-framing in terms of the proportion of fellow supporters who believe that the action has been successful.  From the bottom to top rows, $R$ increases; i.e. the length of time that agents can freely communicate signals to other agents between cycles. Contour plots of different agent types and the DIME variables are available in the Supplementary Materials.


As $\phi$ increases (from left to right columns), latent radicalism begins emerging under a larger space of conditions. For example, for $p = 0.85$ (failure is high), $F = 0.45$ (moderate individual re-framing) and $R = 1$ (short communication; i.e., bottom row, Fig.~\ref{fig:DomActions_Rphi}(g)-~\ref{fig:DomActions_Rphi}(i)), there is low latent radicalism when collective re-framing is high ($\phi = 0.2$) (Fig.~\ref{fig:DomActions_Rphi}(g)), but high latent radicalism when collective re-framing is low $\phi = 0.8$ (Fig.~\ref{fig:DomActions_Rphi}(i)). As $R$ increases, the space of emerging latent radicalism shrinks (as seen by the diminishing green area as we move up the rows of Fig.~\ref{fig:DomActions_Rphi}) since there is more time and capacity for people to interact with other agents. However, the impact of $R$ is weaker compared to that of $\phi$, and $R$ has minimal effects when the threshold for collective re-framing is low (low $\phi$) – it does not matter if fewer protesters are encountered, if those that are encountered primarily see the failure as a success. Indeed, increasing $R$ beyond 10 has minimal effect (see Supplementary Materials). Therefore, latent radicalism emerges especially when it requires many supporters to convince a protester to re-frame failure as success (high $\phi$) and there are limited opportunities to contest it collectively (low $R$).

Active innovation and latent conventionalism are the only other emergent scenarios (Fig.~\ref{fig:DomActions_Rphi}), with comparatively weaker dominance (i.e., the peaks of the teal and blue are much lower than that of the green). The prevalence of active innovator agents grows with $\phi$ and shrinks with $R$, with the opposite behaviour occurring for the space of latent conventionalism. We conclude by noting that these results are robust to different initial conditions (see Supplementary Materials).




\section*{Discussion}
Protest is a highly consequential but complex form of behaviour, shaped by cycles of interactions \cite{snow1992master, tarrow1993cycles} across multiple levels of analysis~\cite{ulug2022conceptual, thomas2022mobilise}. These complexities pose key challenges for the empirical study of collective action and social change \cite{zeineddine2021feeling}. The present research integrates the insights of social psychological models of collective action \cite{louis2020volatility, louis2022failure} with agent-based modelling to address these gaps, examining cycles of interactions within and between protesters and authorities. We explore how complex interactions at lower levels (e.g., within and between protesters, authorities) can shape \textit{emergent} behaviours at the population or movement level in ways that may not be well anticipated in the current literature.

Indeed, our data revealed starkly different emergent outcomes at the two idealised extremes. On the one hand, the models revealed that where the authorities are responsive (low levels of failure) and where people individually (cognitively) and collectively (socially) contest failure, the population was largely conventionally engaged (latent and actively). Stable, sustained low-moderate levels of protest and innovation in partnership with a state which is quick to amend grievances, suggest the paradigm can model a well-functioning democracy~\cite{moghaddam2016_book}.  
 
 On the other hand, in contexts with unresponsive governments (high failure), where such failure is incontrovertible and/or people cannot contest the outcome collectively (e.g., due to repression), protesters generally become inactive but radicalised. It is our contention that widespread latent radicalism – when the population have disengaged but are radicalised – holds the key to understanding how mass protest can emerge seemingly spontaneously (e.g., in Tahrir square). In other words, repressive conditions may set the stage for future radical action because they give rise to the febrile precursor state to widespread revolutionary action~\cite{moghaddam2020psychology} -- the tinder is dry, awaiting a spark. In future extensions of DIMESim, triggering factors such as incidents of pubic injustice and repression \cite{ayanian2021resistance}, disseminated and broadcast allowing for collective re-framing \cite{mcgarty2014technologies} and possibility of resistance, may be included. Such events could create a sudden surge of  commitment (identification), which in the present model would shift actors from inaction to sudden mass action. 

 There is also a clear distinction between the two emergent outcomes in terms of the composition, or heterogeneity, of the movement overall. Where there is a conventional movement, there is a broad mixture of different protester tactics as well as moderate levels of passive (latent) support and overt action. On the other hand, whenever latent radicalism emerges, it occurs in most of the population.

 The model also allows us to examine nonlinear interactions between variables across multiple levels of analysis. Here, the data offer many points of engagement but, for instance, we find that even when the authority consistently signals failure, a population that is sufficiently able to contest failure will continue to act, resulting in a primarily conventionally engaged movement. Repeated movement failure, especially when it sits alongside repression of thought and interaction, is a path to mass latent radicalisation – but the reverse is also true: conventional action dominates where the movements are actually and/or \textit{understood to be}, succeeding in some measure \cite{hornsey2006why}. The psychological “contract” between participants in a movement and authority substantially buffers latent and overt radicalisation (see \cite{drury2000collective, drury2005phenomenology,della2018radicalization}).

 Contemporary theorising of the multi-level nature of collective action identifies myriad other factors that could be incorporated into future models \cite{thomas2022mobilise, zeineddine2021feeling, ulug2022conceptual}. Our model focused on modelling individuals' psychological states within social networks, and was initialised based on a population of people who had already engaged in conventional action (following DIME; \cite{louis2020volatility,louis2022failure}). Future models could theorise and test the mobilisation of those who are yet to act, and include variables common in economics (e.g., cost of action~\cite{makowsky2013agent,acemoglu2008persistence}), political science (e.g., cost of leadership~\cite{lohmann1993signaling}), and sociology (e.g., media framing~\cite{myers2000diffusion}). We also did not model effects in the context of counter-mobilisation where another social movement emerges to oppose the desired change; such interactions would likely produce different emergent dynamics than those documented here~\cite{della2008research,della2015oxford,tarrow1993cycles}. In DIMESim, authorities are represented by a single actor who signals a binary success or failure. Future research could simulate effects with more realism – e.g., authorities could be modelled adapting their signals in response to the protesters' actions, authorities' responses could be constrained by laws and institutional actors~\cite{cantoni2024protests,della2008research}, and the presence of counter-protesters incorporated. 
 Our analysis shows that agent-based models are well-equipped to integrate theoretical, empirical and meta-theoretical insights into a common framework. 

 For decades, scholars have highlighted the importance of avoiding reductionist readings of how people relate to their social world, and seek to change it turn \cite{wright1990responding,saunders2012explaining,tarrow1993cycles,tilly2015social}. People are influenced by and influences on the worlds of which they are a part. The computational approach advanced here provides a promising additional apparatus in the toolkit.

\newpage

 \bibliographystyle{ieeetr} 
 \bibliography{DIME_refs}

\begin{thebibliography}{10}

\bibitem{ortiz2022analysis}
I.~Ortiz, S.~Burke, M.~Berrada, and H.~Saenz~Cort{\'e}s, ``An analysis of world protests 2006--2020,'' {\em World protests: A study of Key protest issues in the 21st century}, pp.~13--81, 2022.

\bibitem{cantoni2024protests}
D.~Cantoni, A.~Kao, D.~Y. Yang, and N.~Yuchtman, ``Protests,'' {\em Annual Review of Economics}, vol.~16, 2024.

\bibitem{roggeband2017handbook}
C.~Roggeband and B.~Klandermans, eds., {\em Handbook of {Social} {Movements} {Across} {Disciplines}}.
\newblock Handbooks of {Sociology} and {Social} {Research}, Cham: Springer International Publishing, 2017.

\bibitem{shuman2023social}
E.~Shuman, A.~Goldenberg, T.~Saguy, E.~Halperin, and M.~van Zomeren, ``{When Are Social Protests Effective?},'' {\em Trends in Cognitive Sciences}, 2023.

\bibitem{paret2021persistent}
M.~Paret, ``The persistent protest cycle: A case study of contained political incorporation,'' {\em Current Sociology}, vol.~69, no.~6, pp.~861--878, 2021.

\bibitem{saunders2012explaining}
C.~Saunders, M.~Grasso, C.~Olcese, E.~Rainsford, and C.~Rootes, ``{Explaining Differential Protest Participation: Novices, Returners, Repeaters, and Stalwarts},'' {\em Mobilization: An International Quarterly}, vol.~17, no.~3, pp.~263--280, 2012.

\bibitem{tilly2015social}
C.~Tilly and L.~J. Wood, {\em Social Movements, 1768-2012}.
\newblock Routledge, 2015.

\bibitem{mccarthy1977resource}
J.~D. McCarthy and M.~N. Zald, ``{Resource Mobilization and Social Movements: A Partial Theory},'' {\em American Journal of Sociology}, vol.~82, no.~6, pp.~1212--1241, 1977.

\bibitem{mcadam1999political}
D.~McAdam, {\em {Political Process and the Development of Black Insurgency, 1930-1970}}.
\newblock University of Chicago Press, 1999.

\bibitem{della2015oxford}
D.~Della~Porta and M.~Diani, eds., {\em {The Oxford Handbook of Social Movements}}.
\newblock Oxford University Press, 2015.

\bibitem{agostini2021toward}
M.~Agostini and M.~van Zomeren, ``Toward a comprehensive and potentially cross-cultural model of why people engage in collective action: A quantitative research synthesis of four motivations and structural constraints,'' {\em Psychological Bulletin}, vol.~147, no.~7, pp.~667–--700, 2021.

\bibitem{becker2017dynamic}
J.~C. Becker and N.~Tausch, ``A dynamic model of engagement in normative and non-normative collective action: Psychological antecedents, consequences, and barriers,'' {\em European Review of Social Psychology}, vol.~26, no.~1, pp.~43--92, 2015.

\bibitem{thomas2022mobilise}
E.~F. Thomas, L.~Duncan, C.~McGarty, W.~R. Louis, and L.~G.~E. Smith, ``Mobilise: A higher-order integration of collective action research to address global challenges,'' {\em Political Psychology}, vol.~43, no.~S1, pp.~107--164, 2022.

\bibitem{tarrow1993cycles}
S.~Tarrow, ``{Cycles of Collective Action: Between Moments of Madness and the Repertoire of Contention},'' {\em Social Science History}, vol.~17, no.~2, pp.~281--307, 1993.

\bibitem{klandermans1984mobilization}
B.~Klandermans, ``Mobilization and participation: Social-psychological expansisons of resource mobilization theory,'' {\em American Sociological Review}, vol.~49, no.~5, pp.~583--600, 1984.

\bibitem{selvanathan2020marches}
H.~P. Selvanathan and J.~Jetten, ``From marches to movements: building and sustaining a social movement following collective action,'' {\em Current Opinion in Psychology}, vol.~35, pp.~81--85, 2020.
\newblock Social Change (Rallies, Riots and Revolutions).

\bibitem{de2013dramatic}
R.~De~La~Sablonni{\`e}re, L.~French~Bourgeois, and M.~Najih, ``Dramatic social change: A social psychological perspective,'' {\em Journal of Social and Political Psychology}, vol.~1, no.~1, 2013.

\bibitem{klandermans1987potentials}
B.~Klandermans and D.~Oegema, ``Potentials, networks, motivations, and barriers: Steps towards participation in social movements,'' {\em American Sociological Review}, vol.~52, no.~4, pp.~519--531, 1987.

\bibitem{cohen2023should}
N.~Cohen-Eick, E.~Shuman, M.~van Zomeren, and E.~Halperin, ``Should i stay or should i go? motives and barriers for sustained collective action toward social change,'' {\em Personality and Social Psychology Bulletin}, p.~01461672231206638, 2023.

\bibitem{louis2020volatility}
W.~Louis, E.~Thomas, C.~McGarty, M.~Lizzio-Wilson, C.~Amiot, and F.~Moghaddam, ``The volatility of collective action: Theoretical analysis and empirical data,'' {\em Political Psychology}, vol.~41, no.~S1, pp.~35--74, 2020.

\bibitem{meyer2004conceptualizing}
D.~S. Meyer and D.~C. Minkoff, ``{Conceptualizing Political Opportunity},'' {\em Social Forces}, vol.~82, no.~4, pp.~1457--1492, 2004.

\bibitem{moskalenko2009measuring}
S.~Moskalenko and C.~McCauley, ``Measuring political mobilization: The distinction between activism and radicalism,'' {\em Terrorism and Political Violence}, vol.~21, no.~2, pp.~239--260, 2009.

\bibitem{mccauley2017understanding}
C.~McCauley and S.~Moskalenko, ``{Understanding Political Radicalization: {The} Two-Pyramids Model},'' {\em American Psychologist}, vol.~72, no.~3, pp.~205--216, 2017.
\newblock Place: US Publisher: American Psychological Association.

\bibitem{della2008research}
D.~Della~Porta, ``{Research on Social Movements and Political Violence},'' {\em Qualitative Sociology}, vol.~31, pp.~221--230, 2008.

\bibitem{wright1990responding}
S.~C. Wright, D.~M. Taylor, and F.~M. Moghaddam, ``Responding to membership in a disadvantaged group: From acceptance to collective protest.,'' {\em Journal of Personality and Social Psychology}, vol.~58, no.~6, p.~994, 1990.

\bibitem{tausch2011radical}
N.~Tausch, J.~Becker, R.~C.~Spears, O.~Christ, R.~Saab, P.~Singh, and R.~N. Siddiqui, ``Explaining radical group behavior: Developing emotion and efficacy routes to normative and nonnormative collective action,'' {\em Journal of Personality and Social Psychology}, vol.~101, no.~1, pp.~129–--148, 2011.

\bibitem{benford2000framing}
R.~D. Benford and D.~A. Snow, ``Framing processes and social movements: An overview and assessment,'' {\em Annual Review of Sociology}, vol.~26, no.~Volume 26, 2000, pp.~611--639, 2000.

\bibitem{snow1992master}
D.~A. Snow and R.~D. Benford, {\em Master frames and cycles of protest}.
\newblock Frontiers in social movement theory, 1992.

\bibitem{louis2022failure}
W.~R. Louis, M.~Lizzio-Wilson, M.~Cibich, C.~McGarty, E.~F. Thomas, C.~E. Amiot, N.~Weber, J.~Rhee, G.~Davies, T.~Rach, {\em et~al.}, ``{Failure Leads Protest Movements to Support More Radical Tactics},'' {\em Social Psychological and Personality Science}, vol.~13, no.~3, pp.~675--687, 2022.

\bibitem{lizzio-wilson2021collective}
M.~Lizzio-Wilson, E.~F. Thomas, W.~R. Louis, B.~Wilcockson, C.~E. Amiot, F.~M. Moghaddam, and C.~McGarty, ``How collective-action failure shapes group heterogeneity and engagement in conventional and radical action over time,'' {\em Psychological Science}, vol.~32, no.~4, pp.~519--535, 2021.
\newblock PMID: 33780273.

\bibitem{Miller2009}
J.~H. Miller and S.~E. Page, {\em Complex Adaptive Systems: An Introduction to Computational Models of Social Life}.
\newblock Princeton University Press, 2009.

\bibitem{ouellette1998habit}
J.~A. Ouellette and W.~Wood, ``Habit and intention in everyday life: The multiple processes by which past behavior predicts future behavior.,'' {\em Psychological Bulletin}, vol.~124, no.~1, p.~54, 1998.

\bibitem{thomas2020testing}
E.~F. Thomas, E.~Zubielevitch, C.~G. Sibley, and D.~Osborne, ``Testing the social identity model of collective action longitudinally and across structurally disadvantaged and advantaged groups,'' {\em Personality and Social Psychology Bulletin}, vol.~46, no.~6, pp.~823--838, 2020.

\bibitem{nepstad2004persistent}
S.~E. Nepstad, ``{Persistent Resistance: Commitment and Community in the Plowshares Movement},'' {\em Social Problems}, vol.~51, no.~1, pp.~43--60, 2004.

\bibitem{becker2014group}
J.~C. Becker and N.~Tausch, ``When group memberships are negative: The concept, measurement, and behavioral implications of psychological disidentification,'' {\em Self and Identity}, vol.~13, no.~3, pp.~294--321, 2014.

\bibitem{della2018radicalization}
D.~Della~Porta, ``{Radicalization: A Relational Perspective},'' {\em Annual Review of Political Science}, vol.~21, no.~1, pp.~461--474, 2018.

\bibitem{jones2008terrorist}
S.~G. Jones and M.~C. Libicki, {\em How terrorist groups end: Lessons for countering al Qa'ida}, vol.~741.
\newblock Rand Corporation, 2008.

\bibitem{skitka2021psychology}
L.~J. Skitka, B.~E. Hanson, G.~S. Morgan, and D.~C. Wisneski, ``The psychology of moral conviction,'' {\em Annual Review of Psychology}, vol.~72, no.~Volume 72, 2021, pp.~347--366, 2021.

\bibitem{hornsey2006why}
M.~J. Hornsey, L.~Blackwood, W.~Louis, K.~Fielding, K.~Mavor, T.~Morton, A.~O'Brien, K.-E. Paasonen, J.~Smith, and K.~M. White, ``Why do people engage in collective action? revisiting the role of perceived effectiveness,'' {\em Journal of Applied Social Psychology}, vol.~36, no.~7, pp.~1701--1722, 2006.

\bibitem{bursztyn2021persistent}
L.~Bursztyn, D.~Cantoni, D.~Y. Yang, N.~Yuchtman, and Y.~J. Zhang, ``{Persistent Political Engagement: Social Interactions and the Dynamics of Protest Movements},'' {\em American Economic Review: Insights}, vol.~3, no.~2, pp.~233--250, 2021.

\bibitem{blackwood2012if}
L.~M. Blackwood and W.~R. Louis, ``If it matters for the group then it matters to me: Collective action outcomes for seasoned activists,'' {\em British Journal of Social Psychology}, vol.~51, no.~1, pp.~72--92, 2012.

\bibitem{thomas2014social}
E.~F. Thomas, C.~McGarty, and W.~Louis, ``Social interaction and psychological pathways to political engagement and extremism,'' {\em European Journal of Social Psychology}, vol.~44, no.~1, pp.~15--22, 2014.

\bibitem{gulliver2021psychology}
R.~Gulliver, S.~Wibisono, K.~S. Fielding, and W.~R. Louis, {\em The Psychology of Effective Activism}.
\newblock Elements in Applied Social Psychology, Cambridge University Press, 2021.

\bibitem{drury2005explaining}
J.~Drury and S.~Reicher, ``Explaining enduring empowerment: a comparative study of collective action and psychological outcomes,'' {\em European Journal of Social Psychology}, vol.~35, no.~1, pp.~35--58, 2005.

\bibitem{granovetter1978threshold}
M.~Granovetter, ``Threshold models of collective behavior,'' {\em American Journal of Sociology}, vol.~83, no.~6, pp.~1420--1443, 1978.

\bibitem{siegel2011does}
D.~A. Siegel, ``{When Does Repression Work? Collective Action in Social Networks},'' {\em The Journal of Politics}, vol.~73, no.~4, pp.~993--1010, 2011.

\bibitem{lohmann1993signaling}
S.~Lohmann, ``A signaling model of informative and manipulative political action,'' {\em American Political Science Review}, vol.~87, no.~2, pp.~319--333, 1993.

\bibitem{kuran1989sparks}
T.~Kuran, ``Sparks and prairie fires: A theory of unanticipated political revolution,'' {\em Public Choice}, vol.~61, no.~1, pp.~41--74, 1989.

\bibitem{acemoglu2008persistence}
D.~Acemoglu and J.~A. Robinson, ``{Persistence of Power, Elites, and Institutions},'' {\em American Economic Review}, vol.~98, no.~1, pp.~267--293, 2008.

\bibitem{rubin2014centralized}
J.~Rubin, ``Centralized institutions and cascades,'' {\em Journal of Comparative Economics}, vol.~42, no.~2, pp.~340--357, 2014.

\bibitem{myers2000diffusion}
D.~J. Myers, ``{The Diffusion of Collective Violence: Infectiousness, Susceptibility, and Mass Media Networks},'' {\em American Journal of Sociology}, vol.~106, no.~1, pp.~173--208, 2000.

\bibitem{centola2013homophily}
D.~M. Centola, ``Homophily, networks, and critical mass: Solving the start-up problem in large group collective action,'' {\em Rationality and Society}, vol.~25, no.~1, pp.~3--40, 2013.

\bibitem{macy2020threshold}
M.~W. Macy and A.~Evtushenko, ``{Threshold models of collective behavior II: The predictability paradox and spontaneous instigation},'' {\em Sociological Science}, vol.~7, no.~26, pp.~628--648, 2020.

\bibitem{epstein2002modeling}
J.~M. Epstein, ``Modeling civil violence: An agent-based computational approach,'' {\em Proceedings of the National Academy of Sciences}, vol.~99, no.~suppl\_3, pp.~7243--7250, 2002.

\bibitem{moro2016understanding}
A.~Moro, ``{Understanding the Dynamics of Violent Political Revolutions in an Agent-Based Framework},'' {\em PLoS One}, vol.~11, no.~4, p.~e0154175, 2016.

\bibitem{ye2021collective}
M.~Ye, L.~Zino, {\v{Z}}.~Mlakar, J.~W. Bolderdijk, H.~Risselada, B.~M. Fennis, and M.~Cao, ``Collective patterns of social diffusion are shaped by individual inertia and trend-seeking,'' {\em Nature Communications}, vol.~12, no.~1, p.~5698, 2021.

\bibitem{makowsky2013agent}
M.~D. Makowsky and J.~Rubin, ``{An Agent-Based Model of Centralized Institutions, Social Network Technology, and Revolution},'' {\em PLoS One}, vol.~8, no.~11, p.~e80380, 2013.

\bibitem{jackson2017agent}
J.~C. Jackson, D.~Rand, K.~Lewis, M.~I. Norton, and K.~Gray, ``Agent-based modeling: A guide for social psychologists,'' {\em Social Psychological and Personality Science}, vol.~8, no.~4, pp.~387--395, 2017.

\bibitem{pierskalla2010protest}
J.~H. Pierskalla, ``Protest, deterrence, and escalation: The strategic calculus of government repression,'' {\em Journal of Conflict Resolution}, vol.~54, no.~1, pp.~117--145, 2010.

\bibitem{moghaddam2020psychology}
F.~M. Moghaddam and M.~J. Hendricks, ``The psychology of revolution,'' {\em Current Opinion in Psychology}, vol.~35, pp.~7--11, 2020.
\newblock Social Change (Rallies, Riots and Revolutions).

\bibitem{drury2005phenomenology}
J.~Drury, C.~Cocking, J.~Beale, C.~Hanson, and F.~Rapley, ``The phenomenology of empowerment in collective action,'' {\em British Journal of Social Psychology}, vol.~44, no.~3, pp.~309--328, 2005.

\bibitem{thomas2019vegetarian}
E.~F. Thomas, S.~M. Bury, W.~R. Louis, C.~E. Amiot, P.~Molenberghs, M.~F. Crane, and J.~Decety, ``Vegetarian, vegan, activist, radical: Using latent profile analysis to examine different forms of support for animal welfare,'' {\em Group Processes \& Intergroup Relations}, vol.~22, no.~6, pp.~836--857, 2019.

\bibitem{pavlovic2024classes}
S.~M. Tomislav~Pavlović and C.~McCauley, ``Two classes of political activists: evidence from surveys of u.s. college students and u.s. prisoners,'' {\em Behavioral Sciences of Terrorism and Political Aggression}, vol.~16, no.~2, pp.~227--247, 2024.

\bibitem{holme2002network}
P.~Holme and B.~J. Kim, ``Growing scale-free networks with tunable clustering,'' {\em Phys. Rev. E}, vol.~65, p.~026107, Jan 2002.

\bibitem{ulug2022conceptual}
{\"O}.~M. Uluğ, M.~Chayinska, and L.~R. Tropp, ``Conceptual, methodological, and contextual challenges in studying collective action: Recommendations for future research,'' {\em TPM: Testing, Psychometrics, Methodology in Applied Psychology}, vol.~29, no.~1, pp.~9--22, 2022.

\bibitem{zeineddine2021feeling}
F.~Bou~Zeineddine and C.~W. Leach, ``Feeling and thought in collective action on social issues: Toward a systems perspective,'' {\em Social and Personality Psychology Compass}, vol.~15, no.~7, p.~e12622, 2021.

\bibitem{moghaddam2016_book}
F.~M. Moghaddam, {\em {The Psychology of Democracy}}.
\newblock American Psychological Association Press, Washington DC, 2016.

\bibitem{ayanian2021resistance}
A.~H. Ayanian, N.~Tausch, Y.~G. Acar, M.~Chayinska, W.-Y. Cheung, and Y.~Lukyanova, ``Resistance in repressive contexts: A comprehensive test of psychological predictors.,'' {\em Journal of Personality and Social Psychology}, vol.~120, no.~4, p.~912, 2021.

\bibitem{mcgarty2014technologies}
C.~McGarty, E.~F. Thomas, G.~Lala, L.~G.~E. Smith, and A.-M. Bliuc, ``New technologies, new identities, and the growth of mass opposition in the arab spring,'' {\em Political Psychology}, vol.~35, no.~6, pp.~725--740, 2014.

\bibitem{drury2000collective}
J.~Drury and S.~Reicher, ``Collective action and psychological change: The emergence of new social identities,'' {\em British Journal of Social Psychology}, vol.~39, no.~4, pp.~579--604, 2000.

\end{thebibliography}

\section*{Acknowledgments}

The work was supported by an Australian Research Council Discovery Project award (DP160101618) to WL, ET and CM. The work of MY was supported by the Western Australian Government, under the Premier's Science Fellowship Program, and by the Australian 

The authors confirm contribution to the paper as follows: study conception and design: ET, MY, WL, SA, CM, SE, MLW; analysis and interpretation of results: MY, SA, TM, ET, WL, LW, SE, CM; draft manuscript preparation: ET, MY, WL, TM, SA. All authors reviewed the results and approved. 

The authors declare no competing interests.

A complete set of the simulation code is available at \url{https://github.com/specialistgeneralist/DIMESim}.

 \newpage

\section*{Materials and Methods}

\subsection*{Mathematical Details of DIMESim Model}


The DIMESim model is an agent-based model, consisting of a population of $n\geq 2$ agents, which we index as agent $i = 1, 2, 3, \hdots, n$. Each agent represents an individual who is aligned with a common movement, and the model unfolds over discrete time-steps $t = 0, 1, 2, \hdots$. At each time step, the model can be broken down into three sequential processes: authority signal and (individual + collective) re-framing, DIME evolution, and updating of actions. We provide details of these in turn, but before doing so, we list the key variables associated with each agent.

\begin{enumerate}
    \item Disidentification, $D_i(t)\in[0,100]$, indicates how disidentified the agent is with the movement.
    \item Innovation, $I_i(t)\in[0,100]$, indicates how willing an agent is to change their action type.
    \item Moralisation, $M_i(t)\in[0,100]$, indicates how moralised the agent is with respect to supporting the movement. 
    \item Energisation, $E_i(t)\in[0,100]$, indicates how energised the agent is to support the movement.
    \item Perceived signal $B_i(t) \in \{-1,1\}$, is whether the agent perceives the protest movement to have succeeded ($-1$) or failed ($1$) at time $t$.
    \item Tactical orientation $x^c_i(t) \in \{-1,1\}$, determines whether the agent is oriented towards radical ($-1$) or conventional ($1$) protest tactics.
    \item Action $x_i(t) \in \{-1,0,1\}$, represents whether the agent is actively participating in the protest using radical tactics ($-1$) or conventional tactics ($1$), or inactive ($0$).
    \item Last active action $x^h_i(t) \in \{-1,1\}$ represents whether the most recent action (before time $t$) taken by agent $i$ was radical ($-1$) or conventional ($1$) action.
\end{enumerate}
The full list of variables and parameters are available in Tables \ref{tab:model_variables} and \ref{tab:model_parameters}. Note that $D_i = 0$ and $D_i = 100$ correspond to the agent being fully identified and fully disidentified with the movement, respectively. Similarly, $M_i = 0$ and $M_i = 100$ represent agent $i$ being completely demoralised and completely moralised with respect to supporting the movement, respectively, while $E_i = 0$ and $E_i = 100$ correspond to an agent $i$ being entirely without energy and completely energised to support the movement, respectively. The innovation term $I_i$ refers to how willing an agent is to change their action type (more details on this in the sequel).

\subsubsection*{Authority signal and re-framing}

At every new time-step $t+1$, the \textit{authority} broadcasts a signal $U(t+1) \in \{-1,1\}$ to all agents, according to the following equation: 
\begin{equation}\label{eq:global_broadcast}
    U(t+1) = 
    \begin{cases}
        1, & \text{with probability } p \\
        -1, & \text{with probability } 1-p 
    \end{cases},
\end{equation}
where $U(t+1) = -1$ and $U(t+1) = 1$ is a signal of `success' and `failure', respectively, and $p\in[0,1]$ is the probability of the authority signalling `failure'. 

If the authority signal is `success', all agents accept it. But if it is `failure', then each agent contests the signal, re-framing it as $B_i^{IR}(t+1) \in \{-1,1\}$ according to the following equation:
\begin{equation}\label{eq:perceived_broadcast_IR}
    B_i^{IR}(t+1) = 
    \begin{cases}
        -1, & \text{if } U(t+1) = -1 \text{ or }\frac{[100-D_i(t)]Z_i(t)}{100}>F \\
        1, & \text{otherwise}        
    \end{cases}
\end{equation}
where, $D_i(t) \in [0,100]$ is the disidentification level of the agent, $F\in [0,1]$ is a threshold parameter, and $Z_i(t) \in [0,1]$ is a random number that is at every time step $t$ and for each agent sampled from a uniform distribution on the interval $[0,1]$. We have assumed the threshold $F$ to be homogeneous among the agents, although individual values can be chosen if the population is strongly heterogeneous. Therefore, agent~$i$'s probability of re-framing failure to success (when $U(t+1) = 1$) increases as disidentification, $D_i(t)$, decreases, and as the parameter $F$ decreases. 

If the agent still perceives the signal as `failure' after individual re-framing (IR), then $R\geq 1$ rounds of collective re-framing (CR) happen, after which the final perceived signal $B_i(t+1) \in \{-1,1\}$ is given by the following equation:
\begin{equation}\label{eq:perceived_broadcast_CR}
    B_i(t+1):= Q^R_i (t+1)
\end{equation}
where $Q^j_i(t+1) \in \{-1,1\}$ denotes the signal perceived by agent $i$ after the $j^{th}$ CR round. 

The quantity $Q^R_i(t+1)$ is defined recursively as follows. We define $Q^0_i(t+1):=B^{IR}_i(t+1)$. Now, supposing that $Q^j_i(t+1)$ has been defined for some integer $0\leq j<R$, we define
\begin{equation}\label{eq:perceived_broadcast_CR_intermediate}
    Q^{j+1}_i(t+1) := 
    \begin{cases}
        -1, & S^{j}_i(t+1)>\phi \\
        Q^j_i(t+1), & \text{otherwise}      
    \end{cases}
\end{equation}
where $S^j_i(t+1) \in[0,1]$ is defined in \eqref{eq:proportion_success_neighbourhood} below. In particular, $S^j_i(t+1)$ denotes the proportion of the neighbours of agent $i$ who perceive the authority signal as `success' at the $j$th CR round of time-step $t+1$, and $\phi\in[0,1]$ is a threshold parameter. We assume that threshold $\phi$ is homogeneous among the agents. 

We represent connections between agents using an undirected unweighted graph $G=(V, E)$ with node set $V =\{1,2,\dots, n\}$ and edge set $E\subseteq V\times V$. Each node $i\in V$ denotes an agent, and an undirected edge $(i,j)\in E$ connects agent $i$ to agent $j$ if the two agents `interact' with one another. Here, an interaction means the ability for the two agents to share their $Q^j_i$ values, i.e., to discuss their perceptions of the authority signal. Note that we assume an agent does not interact with themselves, i.e., $(i,i)\notin E$ for all $i\in V$. The neighbourhood of agent $i$, denoted by $\mathcal{N}_i$, is defined to be the set $\mathcal{N}_i:=\{j\in V : (i,j)\in E\}$, and agent $j$ is said to be a neighbour of agent $i$ if and only if $j\in \mathcal{N}_i$. The proportion of neighbours of agent $i$ who perceive the authority signal as success at the $j$th stage of CR, $S_i^j(t+1)$, is thus defined as,
\begin{equation} \label{eq:proportion_success_neighbourhood}
    S_i^j(t+1) =
    \begin{cases}
        \frac{1}{2|\mathcal{N}_i|} \sum_{k\in\mathcal{N}_i} 1-Q^j_k(t+1), & \mathcal{N}_i\ne \emptyset \\
        0, & \mathcal{N}_i=\emptyset
    \end{cases}
\end{equation}
where $|\cdot |$ denotes set cardinality. Note that in the case where agent $i$ has no neighbours, \eqref{eq:perceived_broadcast_CR} assigns $B^{IR}_i(t+1)$ to the final perceived signal value $B_i(t+1)$ since $0\not > \phi$ for any $\phi\in [0,1]$.

We note that when $U(t+1) = -1$, then the sequence of individual and collective re-framing, viz. \eqref{eq:perceived_broadcast_IR} through \eqref{eq:proportion_success_neighbourhood}, always result in $B_i(t+1) = -1$ for all agents; i.e. when the authority signal is success, the perceived signal by all agents is always success. Moreover, re-framing only allows conversion of failure signals to success, so that if $B_i^{IR}(t+1) = -1$ (agent~$i$ successfully re-frames failure to success at the individual re-framing stage), then the collective re-framing stage cannot change this outcome: $Q_i^j(t+1) = -1$ for all $j =  0, 1, \hdots, R$.

\subsubsection*{DIME evolution}

At the end of the collective re-framing stage, each agent has their own understanding of whether the protest movement was successful at time $t+1$, viz. $B_i(t+1)$. The DIME theoretical model proposes a framework for how a protester's internal psychological (DIME) states respond to the success or failure of a movement. Incorporating the full meta-analysis results~\cite[Table~2]{louis2022failure}, each agent~$i$ updates their DIME states according to the following four regression equations:
\begin{subequations}\label{eq:DIME_update}
    \begin{align}
        D_i(t+1) & = P\left(D_i(t) + \beta_D B_i(t+1) + \lambda_D x^c_i(t) + \gamma_D B_i(t+1) x^c_i(t) + \omega_i^D(t)\right),  \label{eq:D_update_v2} \\
        I_i(t+1) & = P\left(I_i(t) + \beta_I B_i(t+1) + \lambda_I x^c_i(t) + \gamma_I B_i(t+1) x^c_i(t) + \omega_i^I(t)\right),  \label{eq:I_update_v2}\\
        M_i(t+1) & = P\left(M_i(t) + \beta_M B_i(t+1) + \lambda_M x^c_i(t) + \gamma_M B_i(t+1) x^c_i(t) + \omega_i^M(t)\right),   \label{eq:M_update_v2} \\
        E_i(t+1) & = P\left(E_i(t) + \beta_E B_i(t+1) + \lambda_E x^c_i(t) + \gamma_E B_i(t+1) x^c_i(t) + \omega_i^E(t)\right).  \label{eq:E_update_v2}
    \end{align}
\end{subequations}
Here, $\beta_D$, $\beta_I$, $\beta_M$, $\beta_E$ are the aggregate partial correlations adapted from the first column (Protest Outcome) of \cite[Table~2]{louis2022failure}, and they describe the gradient or velocity of change in the DIME states based on the (perceived) protest outcome $B_i(t+1)$. The $\lambda$ and $\gamma$ values are adapted from the coefficients in the second and third columns of \cite[Table~2]{louis2022failure}, that describe the gradient based on the orientation of the agent, $x^c_i(t) \in \{-1, 1\}$, and the product $B_i(t+1) x^c_i(t)$, respectively. The terms $\omega_i(t)$ are random noise with uniform distribution over the unit interval $[0,1]$. We impose a saturation so that the four DIME states cannot be greater than $100$ or less than~$0$ via $P:\mathbb{R}\to[0,100], x\mapsto P(x)$ defined as,
\begin{equation}\label{eq:P_saturation}
    P(x) = 
    \begin{cases}
        0, & x<0 \\
        x, & 0\leq x\leq 100 \\
        100, & x>100
    \end{cases}
\end{equation}

\subsubsection*{Decision to act and how to act}
After updating their internal psychological state, each agent decides whether they want to take action in the current time step, $A_i(t+1) \in \{0, 1\}$, and whether they want to stick with their previous orientation (conventional or radical) or innovate to the alternative, $C_i(t+1) \in \{-1, 1\}$, according to the following equation:
\begin{subequations}\label{eq:CA_update}
\begin{align}
    A_i(t+1) & = 
    \begin{cases}
        0, & \text{if } D_i(t+1) > \frac{I_i(t+1) + M_i(t+1) + E_i(t+1)}{3}\\
        1 & \text{otherwise}
    \end{cases} \label{eq:A_update}\\
    C_i(t+1) & =
    \begin{cases}
        -1, & \text{if }  I_i(t+1) > \frac{M_i(t+1) + E_i(t+1)}{2} \\
        1 & \text{otherwise}
    \end{cases} \label{eq:C_update}
\end{align}
\end{subequations}
where $A_i(t+1) = 0$ and $A_i(t+1) = 1$ refer to the agent being latent (inactive) and active, respectively, and $C_i(t+1) = -1$ and $C_i(t+1) = 1$ refer to the agent innovating and sticking with their previous orientation, respectively

The orientation $x^c_i(t) \in \{-1, 1\}$ and action $x_i(t) \in \{-1, 0, 1\}$ that each agent takes is then calculated as:
\begin{subequations}\label{eq:x_update}
    \begin{align}
        x^c_i(t+1) &= C_i(t+1) x_i^h(t + 1) \\
        x_i(t+1) &= A_i(t+1) x^c_i(t+1),
    \end{align}
\end{subequations}
where $x^c_i(t) = -1$ and $1$ indicate the agent is oriented towards radical tactics and conventional tactics respectively. Then, $x_i(t) = -1, 0$ and $1$ indicate that the agent is taking radical action, no action, and conventional action respectively.

Note that $x_i^h(t) \in \{-1,1\}$ is the last active action of agent~$i$, and not the orientation. I.e., the most recent conventional or radical action taken by agent $i$ before the current time-step. Define $x^h_i(0):=1$ (conventional action). Now, supposing inductively that $x^h_i(t)$ has been defined for $t\geq 0$, we define
\begin{equation}\label{eq:xh_update}
    x^h_i(t + 1):= \begin{cases}        
        x_i(t), & \text{if } x_i(t) \neq 0 \\
        x^h_i(t), & \text{otherwise}        
    \end{cases}
\end{equation}

The determination of the variables $x_i(t+1)$ and $x^c_i(t+1)$ is the final event for the new time step $t+1$, determining whether the agent is active (and what tactics the agent is oriented towards). The next iteration then begins with the authority signalling success/failure, as described above in \eqref{eq:global_broadcast}.

\begin{table}[h!]
    \centering
    \caption{Model variables}
    \begin{tabular}{c  c  p{0.8\textwidth}} 
        Variable & Range & Description (with respect to agent~$i$, for $i\in \{1,2,\hdots, n\}$) \\ \midrule
        $\omega_i(t)$ & $[0,1]$ & Random noise with uniform distribution. \\
        $A_i(t)$ & $\{0, 1\}$ & Decision to act: Whether the agent chooses to take action ($1$) or not ($0$) at time $t$. \\
        $B_i(t)$ & $\{-1,1\}$ & Final perceived signal: Whether the agent perceives the protest movement to have succeeded ($-1$) or failed ($1$) at time $t$ after attempting individual and collective re-framing. \\
        $B^{IR}_i(t)$ & $\{-1,1\}$ & Perceived signal after individual re-framing. \\
        $C_i(t)$ & $\{-1, 1\}$ & Decision to innovate: Whether the agent chooses, at time $t$, to innovate to the alternative ($-1$) or stick with their previous choice of tactic ($1$) \\
        $D_i(t)$ & $[0,100]$ & Disidentification: Indicates how disidentified the agent is with the movement. \\
        $E_i(t)$ & $[0,100]$ & Energisation: Indicates how energised the agent is to support the movement. \\
        $I_i(t)$ & $[0,100]$ & Innovation: Indicates how willing an agent is to change their action type. \\
        $M_i(t)$ & $[0,100]$ & Moralisation: Indicates how moralised the agent is with respect to supporting the movement. \\
        $Q^{j}_i(t)$ & $\{-1,1\}$ & Perceived signal after the $j^{th}$ collective re-framing round. \\
        $S^j_i(t+1)$ & $[0,1]$ & The proportion of the neighbours of agent $i$ who currently perceive the signal as `success' at the $j$th CR round of time-step $t+1$ \\
        $U(t)$ & $\{-1, 1\}$ & Authority signal: The signal by the authority \\
        $x_i(t)$ & $\{-1,0,1\}$ & Action: Represents whether the agent is actively participating in the protest using radical tactics ($-1$) or conventional tactics ($1$), or inactive ($0$). \\
        $x^c_i(t)$ & $\{-1,1\}$ & Tactical orientation: Determines whether the agent is oriented towards radical ($-1$) or conventional ($1$) protest tactics. \\
        $x^h_i(t)$ & $\{-1,1\}$ & Last active action: Whether the most recent action (before time $t$) taken by agent $i$ was radical ($-1$) or conventional ($1$) action. \\
        $Z_i(t)$ & $[0, 1]$ & Random number generated at every time step, as part of the individual re-framing process
    \end{tabular}    
    \label{tab:model_variables}
\end{table}

\begin{table}[h!]
    \centering
    \caption{Model parameters}
    \begin{tabular}{c  c  p{0.8\textwidth}} 
        Parameter & Range & Description \\ \midrule
        $\beta$ & $\mathbb{R}$ & DIME gradient based on the (perceived) protest outcome. \\
        $\gamma$ & $\mathbb{R}$ & DIME gradient based on the interaction between perceived protest outcome and agent orientation. \\
        $\lambda$ & $\mathbb{R}$ & DIME gradient based on the agent orientation. \\
        $\phi$ & $[0, 1]$ & Threshold for collective re-framing: As $\phi$ increases, it becomes more difficult to collectively re-frame failure as a success.  An agent requires a greater fraction of its network neighbours to be perceiving success, in order for the agent to also perceive success. \\
        $F$ & $[0 , 1]$ & Threshold for individual re-framing: As $F$ increases, it becomes more difficult to individually re-frame failure as a success. \\
        $n$ & $\geq 2$ & Number of agents \\
        $p$ & $[0, 1]$ & Probability of failure signal: Low values indicate that the movement is succeeding, while high values indicate that the movement is failing. \\
        $R$ & $\geq 1$ & Length of free communication: As $R$ increases, protesters are able to more freely communicate with each other during the collective re-framing stage. \\
        $T$ & $\geq 1$ & Number of total time-steps in the simulation
    \end{tabular}    
    \label{tab:model_parameters}
\end{table}

\subsection*{Simulation setup}

Here, we provide the details on the simulation process used to generate the main results of our study, described in the main article. A complete set of the simulation code is available on \url{https://github.com/specialistgeneralist/DIMESim}.

For each setting of model parameters ($p, F, \phi, R$), we performed 20 independent runs of our simulations with 1000 agents ($n = 1,000$) to 10,000 time steps ($T = 10,000$). For initial conditions, we assumed all agents began as active, with conventional tactics: $A_i(0) = 1, C_i(0) = 1, x^h_i(0) = 1 \; \forall \; i = 1, 2, \dots, n$. The initial DIME variables ($D_i(0)$, $I_i(0)$, $M_i(0)$, and $E_i(0)$) and the gradients ($\beta$, $\lambda$, and $\gamma$) were sampled from normal distributions with means and standard deviations taken from  \cite{louis2022failure}, as shown in Table \ref{tab:DIME_parameters}.
\begin{table}[h!]
    \centering
    \caption{Mean and standard deviations of normal distributions sampled from when generating initial values (adapted from \cite[Table~1]{louis2022failure}) and gradients (adapted from \cite[Table~2]{louis2022failure}) of DIME variables.}
    \begin{tabular}{c *{4}{| c c}} 
         Variable & \multicolumn{2}{|c}{Initial Value} & \multicolumn{2}{|c}{$\beta$} & \multicolumn{2}{|c}{$\lambda$} & \multicolumn{2}{|c}{$\gamma$} \\
         & Mean & Std. Dev. & Mean & Std. Dev. & Mean & Std. Dev. & Mean & Std. Dev. \\\midrule  
         $D$ & 25 & 20 & 2.33 & 1 & -7.33 & 1 & -0.67 & 1 \\
         $I$ & 16.67 & 30 & 0 & 0.67 & 0.33 & 0.67 & 1.67 & 0.67 \\
         $M$ & 58.33 & 21.67 & 1.33 & 1 & -0.33 & 1 & 0.33 & 1 \\
         $E$ & 66.67 & 15 & 3.33 & 0.67 & 1.67 & 0.67 & 1.67 & 0.67 \\
    \end{tabular}    
    \label{tab:DIME_parameters}
\end{table}

Each data point in the timeseries plots is calculated as the rolling average of the corresponding property over the previous 20 time steps and over the independent runs. Each data point in the heatmap, contour, and bar charts is calculated as the temporal average of the last 500 time steps in the simulation and over the independent runs. For actions, the values represent the percentage of the population that holds the corresponding characteristic. For DIME, the values represent the average value among all agents. To determine the population proportions for the different action types when reporting results, the following classification scheme is used,
\begin{itemize}
    \item Active conventional agent: $A_i(t) = 1; \quad C_i(t) = 1; \quad x^h_i(t) = 1$
    \item Active innovator: $A_i(t) = 1;\quad C_i(t) = -1;\quad x^h_i(t) = -1, 1$
    \item Active radical agent: $A_i(t) = 1;\quad C_i(t) = 1;\quad x^h_i(t) = -1$
    \item Latent conventional agent: $A_i(t) = 0;\quad C_i(t) = 1;\quad x^h_i(t) = 1$
    \item Latent innovator: $A_i(t) = 0;\quad C_i(t) = -1;\quad x^h_i(t) = -1, 1$
    \item Latent radical agent: $A_i(t) = 0;\quad C_i(t) = 1;\quad x^h_i(t) = -1$
\end{itemize}



The networks used for the collective re-interpretation process were synthetic random networks generated using the Holme-Kim algorithm~\cite{holme2002network}, which has several network characteristics of real-world social networks, as a power-law degree distribution, high clustering coefficient, and short characteristic path length. For each randomly generated network, the Holme-Kim algorithm starts with a complete subgraph of 13 nodes ($N_0$). Then, nodes are added one-by-one, and each newly added node is connected to the existing nodes via 6 edges ($m$) with, on average, 5 of these being created by a triad-formation step ($m_t$) and 1 being created by preferential attachment. See the Supplementary Materials for an example of a Holme-Kim network compared with the Erdős–Rényi network of similar characteristics.

To summarise, the algorithm for the simulations is as follows:
\begin{enumerate}
    \item Generate the network graph $G$ using the Holme-Kim algorithm.
    \item Sample the stochastic parameters ($\beta, \lambda, \gamma$) using the distribution parameters given in Table \ref{tab:DIME_parameters}
    \item Set the initial value ($t = 0$) for all variables
    \begin{enumerate}
        \item Set $A_i(0) = 1, C_i(0) = 1, x^h_i(0) = 1 \; \forall \; i = 1, 2, \dots, n$
        \item $D_i(0), I_i(0), M_i(0), E_i(0)$ are sampled using the distribution parameters given in Table \ref{tab:DIME_parameters}, and applying the saturation function, \eqref{eq:P_saturation}, to ensure all values lie in $[0,100]$
        \item All other variables are set the default initial value (0), as they are either not referenced in the simulation or are overwritten with a calculation before their first reference
    \end{enumerate}
    \item Run through time-steps $t=1, 2, \dots, T$. At each time step, run through \eqref{eq:global_broadcast} -- \eqref{eq:xh_update}. 
\end{enumerate} 

\setcounter{figure}{0}
\renewcommand\thefigure{S\arabic{figure}}
\setcounter{table}{0}
\renewcommand\thetable{S\arabic{table}}
\numberwithin{equation}{section}
\renewcommand\theequation{\thesection\arabic{equation}}
\ifx\IncludeSupplementary\undefined 
    \refstepcounter{figure}\label{fig:ParameterSweep_p-F_DIME}
    \refstepcounter{figure}\label{fig:ParameterSweep_nu-R_p-F}
    \refstepcounter{figure}\label{fig:ParameterSweep_nu-R_Action}
    \refstepcounter{figure}\label{fig:ParameterSweep_nu-R_DIME}
    \refstepcounter{figure}\label{fig:InitialActions}
    \refstepcounter{figure}\label{fig:NetworkComparison}
    
    \refstepcounter{table}\label{tab:p_F_regions}
    \refstepcounter{table}\label{tab:initial_conditions}
\else 
    \clearpage
    
    \begin{center}
        \Huge Supplementary Materials 
        
        \LARGE {Repeated and incontrovertible collective action failure leads to protester disengagement and radicalisation }

        
    \end{center}

\tableofcontents

\section{Sensitivity of DIME Values to $p$ and $F$}
In the main text, we presented and discussed in Fig.~\ref{fig:ParameterSweep_pF} and \ref{fig:DominantActions_pF} the sensitivity of the long-run steady state population fraction of the six individual agent types as well as the dominant agent type, that emerge from our simulations when varying parameters $p$ and $F$ while keeping parameters $\phi = 0.8$ and $R=10$ constant. Here, we supplement that discussion by presenting and discussing the sensitivity of the average DIME values among agents (Fig.~\ref{fig:ParameterSweep_p-F_DIME}a -- \ref{fig:ParameterSweep_p-F_DIME}d).

\begin{figure}[H]
	\centering \includegraphics[width=\textwidth]{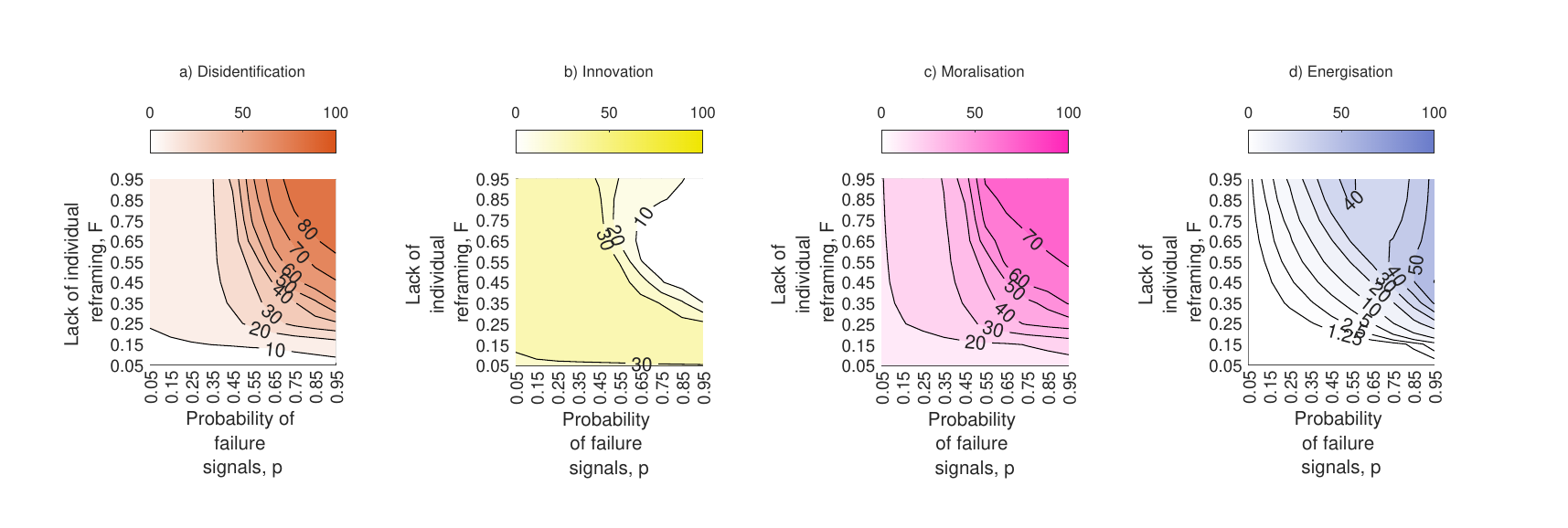}
	\caption{Sensitivity of the long-run steady state average DIME values among agents when varying $p$ and $F$ for constant $\phi = 0.8$ and $R=10$.} \label{fig:ParameterSweep_p-F_DIME}
\end{figure}

We observe that disidentification and moralisation vary from very low to very high, energisation from very low to medium, and innovation remains low throughout this parameter space. Comparing with Fig.~\ref{fig:ParameterSweep_pF}, we observe that the contours of DIME values roughly match the contours for the population fractions, especially for active conventional protesters, latent conventional protesters and latent radical protesters. In particular, as $p$ and $F$ increase: 1) disidentification, moralisation, and energisation increase in value; 2) innovation weakly decreases; 3) population fractions of active conventional protesters and latent conventional protesters decrease; 4) population fraction of latent radical protesters increase; and 5) population fraction of active innovators weakly decrease. Hence, looking at Fig.~\ref{fig:DominantActions_pF}, strong presence of latent radical protesters in a population is associated with high average values of disidentification and moralisation, medium average values of energisation, and low average values of innovation among the population. Meanwhile, moderate presence of active conventional protesters, latent conventional protesters, and active innovators in a population is associated with low average values of disidentification, moralisation, and energisation, and moderate average values of innovation among the population. This dual scenario behaviour matches the results and discussion of Fig.~\ref{fig:NetworkTimeSeriesComposite}.

\newpage
\section{Sensitivity of Population Fractions of Protester Types and DIME Values to $p, F, \phi$ and $R$}
In the main text, we presented and discussed in Fig.~\ref{fig:DomActions_Rphi} the sensitivity of the long-run steady state agent type and population fraction of the dominant agent type that emerge from our simulations for various values of the four primary parameters, $p, F, \phi,$ and $R$. Here, we supplement that discussion by presenting and discussing the sensitivity of the population fractions of the six individual agent types (Fig.~\ref{fig:ParameterSweep_nu-R_p-F_ConHab} -- \ref{fig:ParameterSweep_nu-R_p-F_InaRadHab}), and the average DIME values among agents (Fig.~\ref{fig:ParameterSweep_nu-R_p-F_D} -- \ref{fig:ParameterSweep_nu-R_p-F_E}). Figure~\ref{fig:ParameterSweep_nu-R_p-F_DominantAction} is a flattened view of Fig.~\ref{fig:DomActions_Rphi}, where it is easier to identify the exact regions that each protestor type holds dominant. Note that Fig.~\ref{fig:ParameterSweep_pF}, \ref{fig:DominantActions_pF} and \ref{fig:ParameterSweep_p-F_DIME} are also present within Fig.~\ref{fig:DomActions_Rphi} and \ref{fig:ParameterSweep_nu-R_p-F} as the top-right corner subplots ($\phi=0.8$ and $R=10$).
\begin{figure}[H]
	\centering
		
	\subfloat[Active Conventional Agents]{
		\centering \includegraphics[width = 0.5 \textwidth]{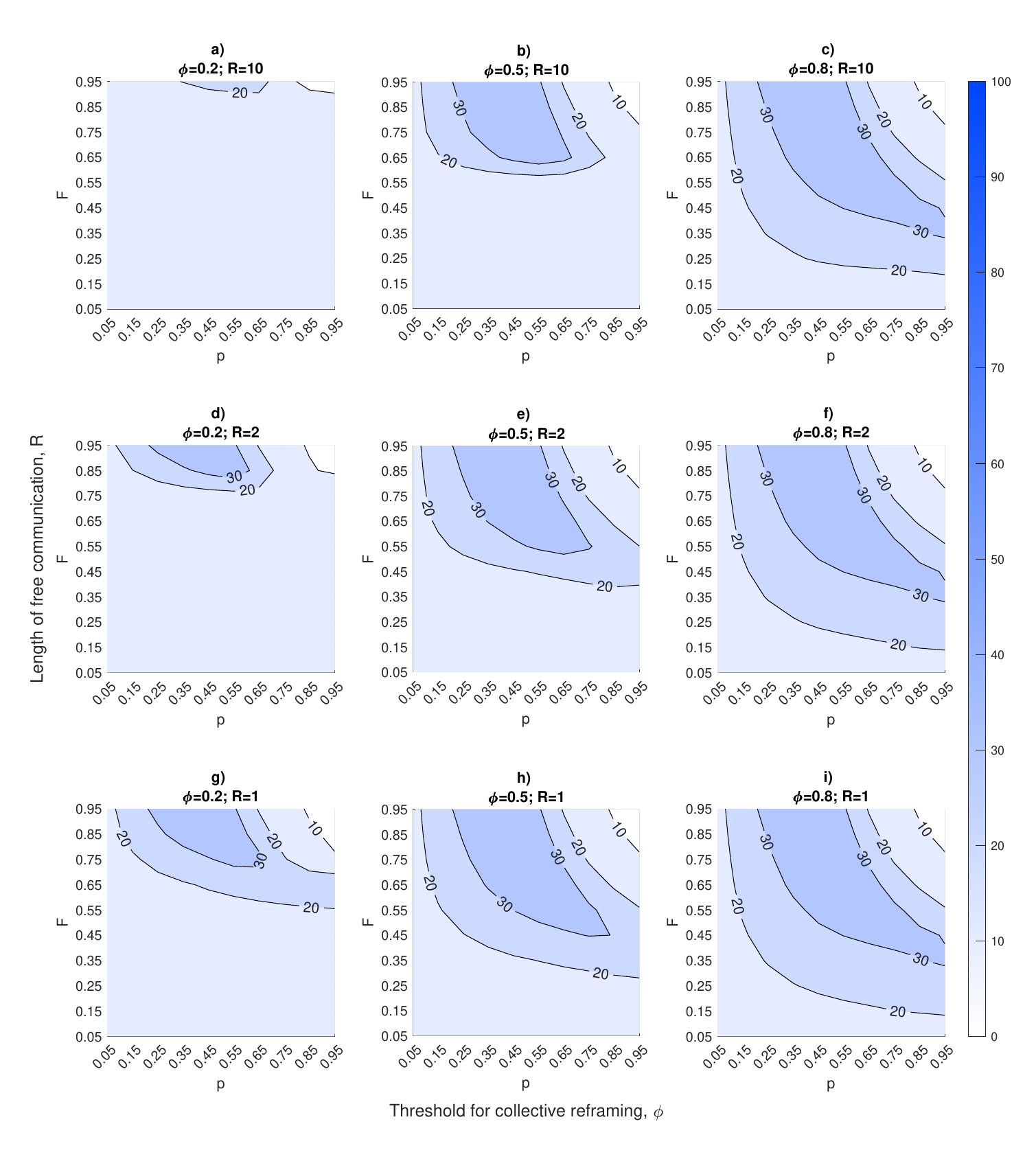}
		\label{fig:ParameterSweep_nu-R_p-F_ConHab}} 
    \subfloat[Latent Conventional Agents]{
		\centering \includegraphics[width = 0.5 \textwidth]{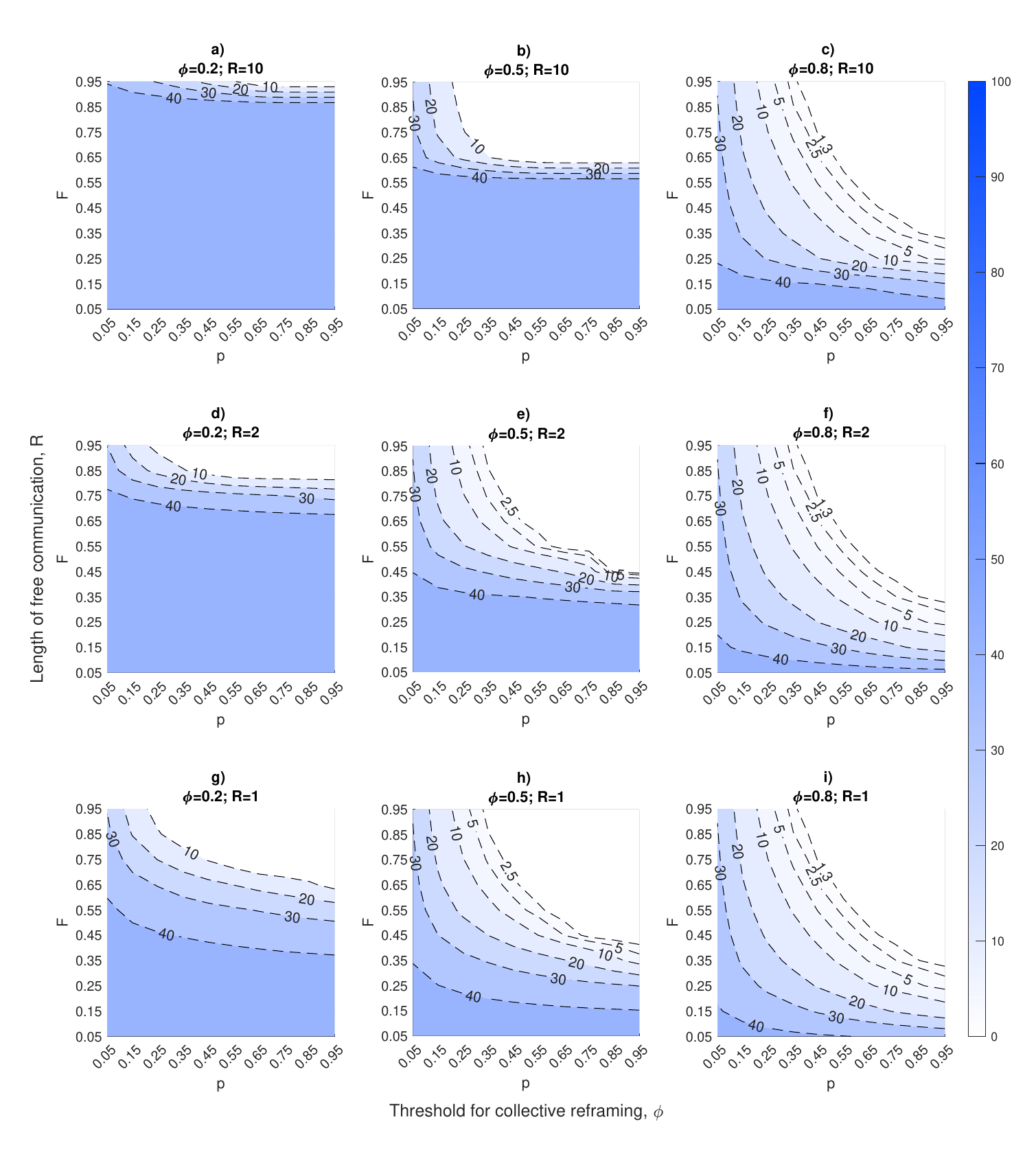}
		\label{fig:ParameterSweep_nu-R_p-F_InaConHab}}

    \subfloat[Active Innovators]{
		\centering \includegraphics[width = 0.5 \textwidth]{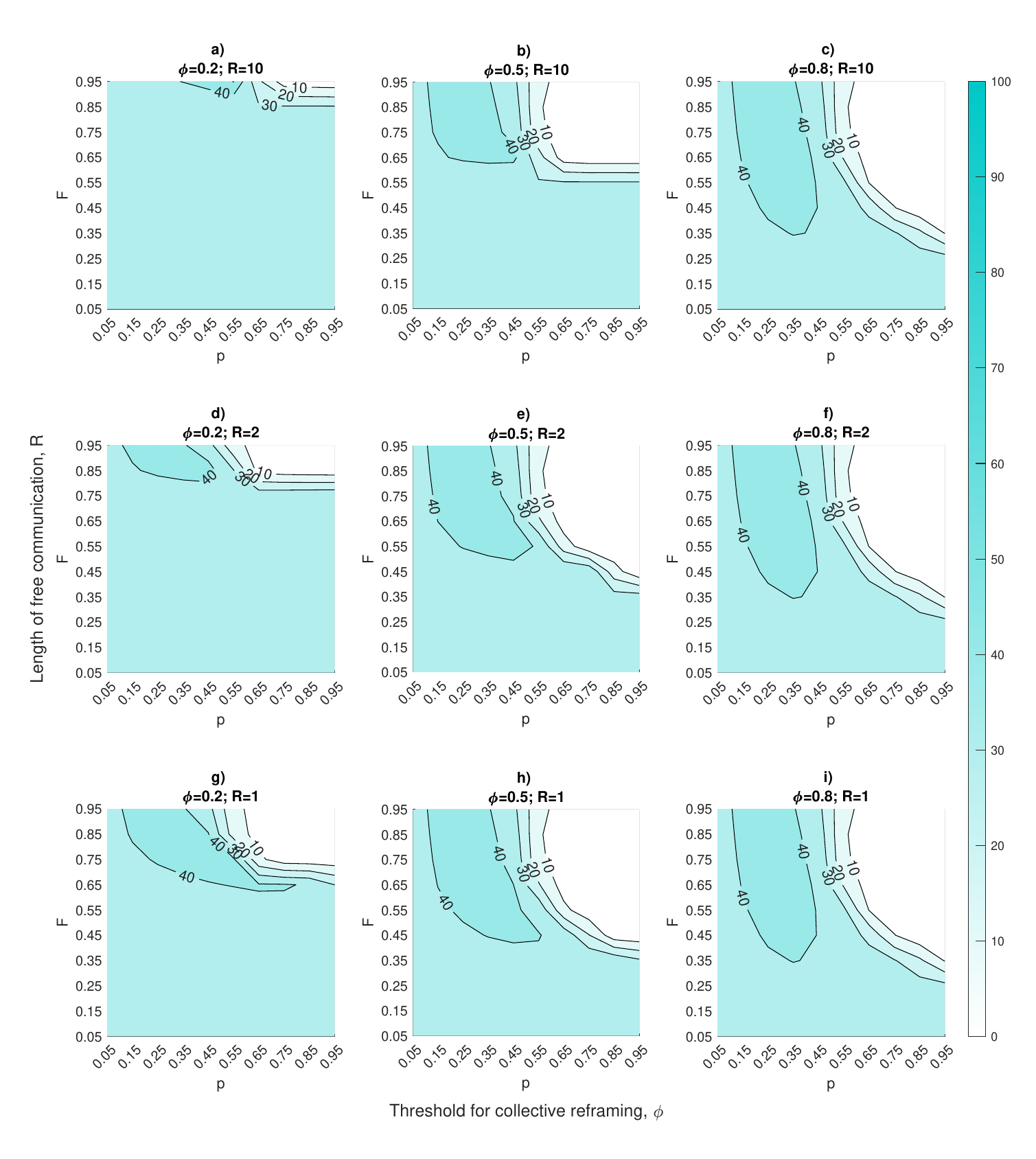}
		\label{fig:ParameterSweep_nu-R_p-F_Inv}} 
    \subfloat[Latent Innovators]{
		\centering \includegraphics[width = 0.5 \textwidth]{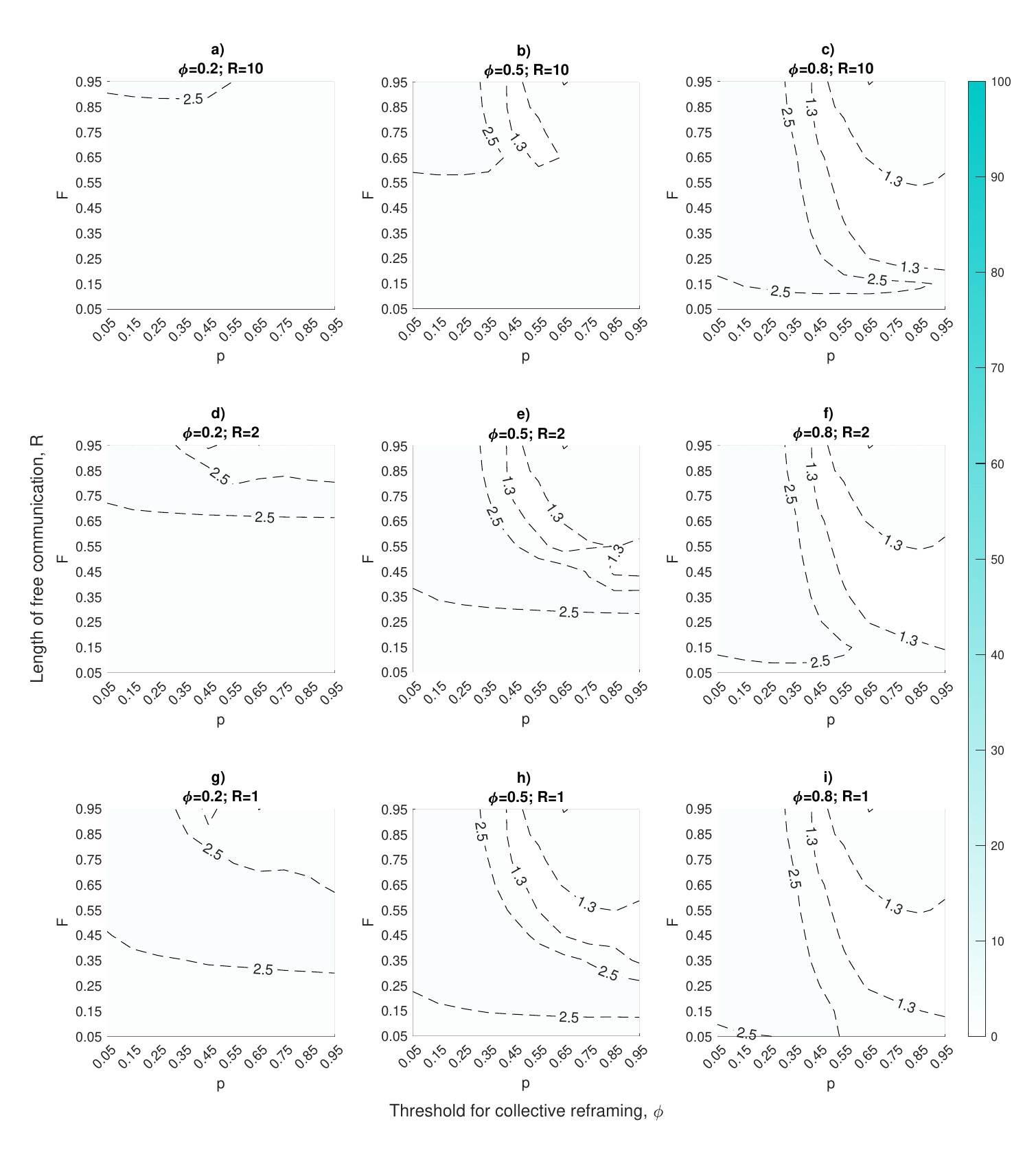}
        \label{fig:ParameterSweep_nu-R_p-F_InaInv}}
\end{figure}

\begin{figure}[H]
    \centering
    \subfloat[Active Radical Agents]{
		\centering \includegraphics[width = 0.5 \textwidth]{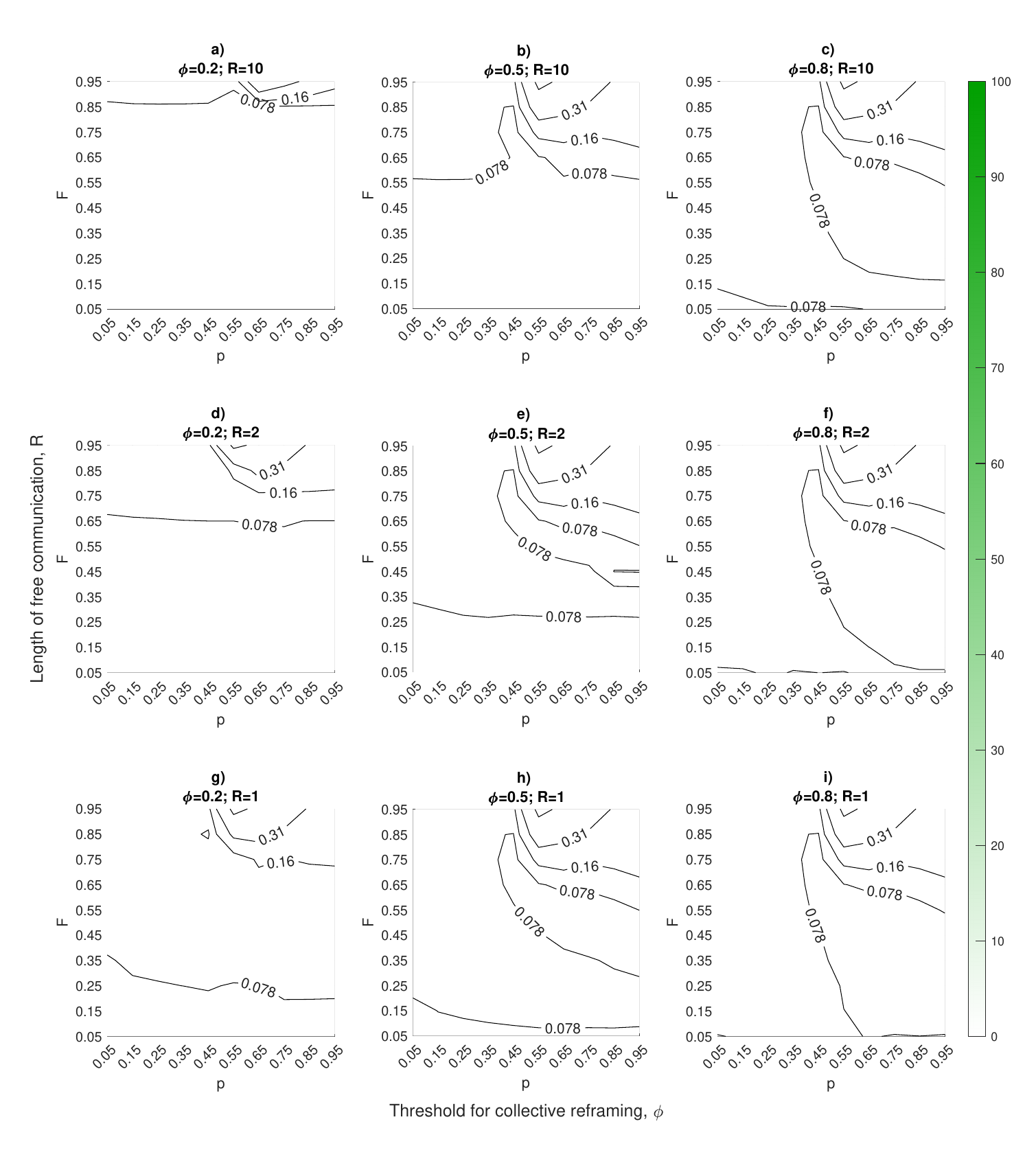}
		\label{fig:ParameterSweep_nu-R_p-F_RadHab}} 
    \subfloat[Latent Radical Agents]{
		\centering \includegraphics[width = 0.5 \textwidth]{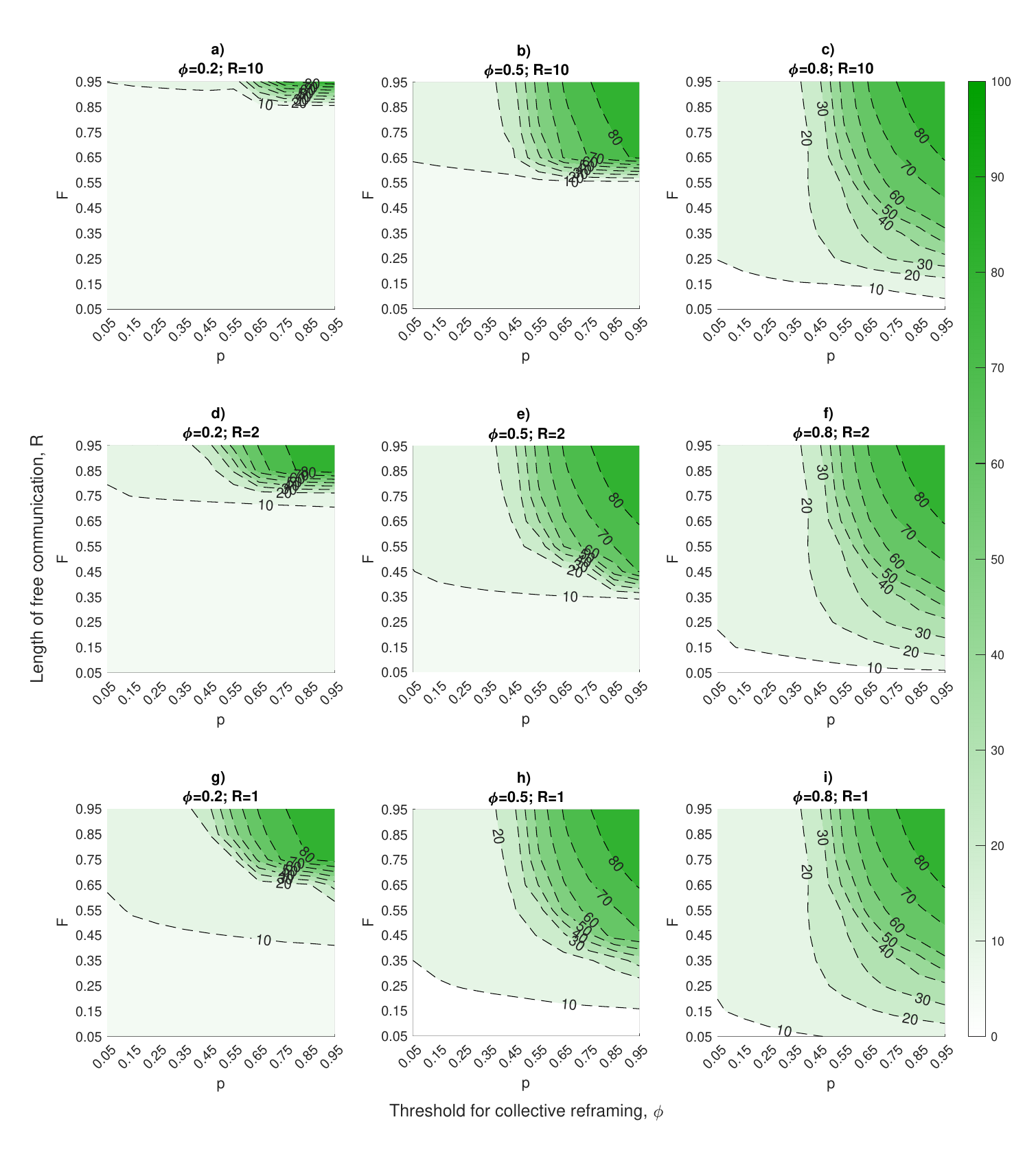}
        \label{fig:ParameterSweep_nu-R_p-F_InaRadHab}}

    \subfloat[Dominant Action]{
		\centering \includegraphics[width = 0.8 \textwidth]{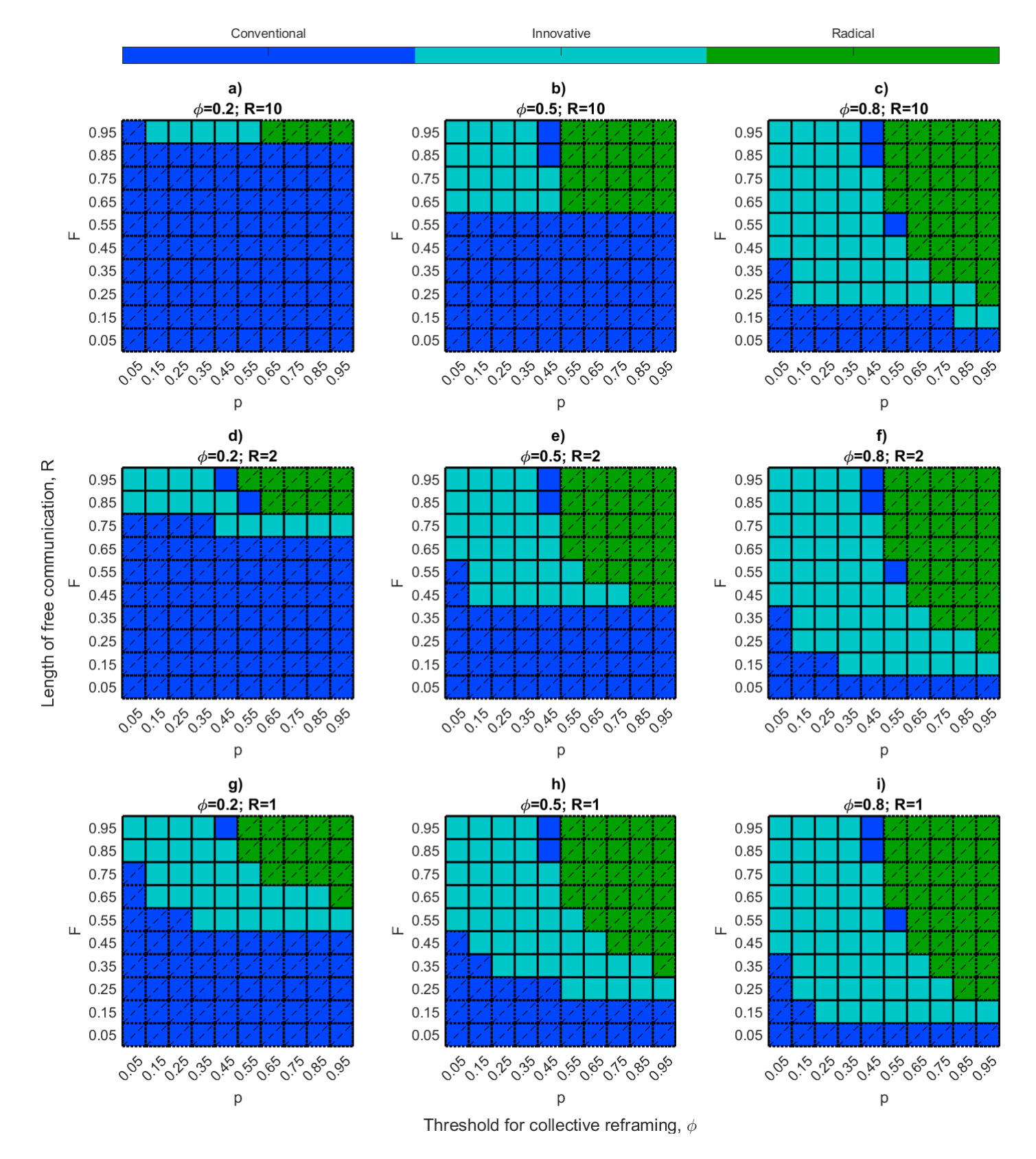}
        \label{fig:ParameterSweep_nu-R_p-F_DominantAction}}
\end{figure}
		
\begin{figure}[H]
	\centering
	\subfloat[Disidentification]{
		\centering \includegraphics[width = 0.5 \textwidth]{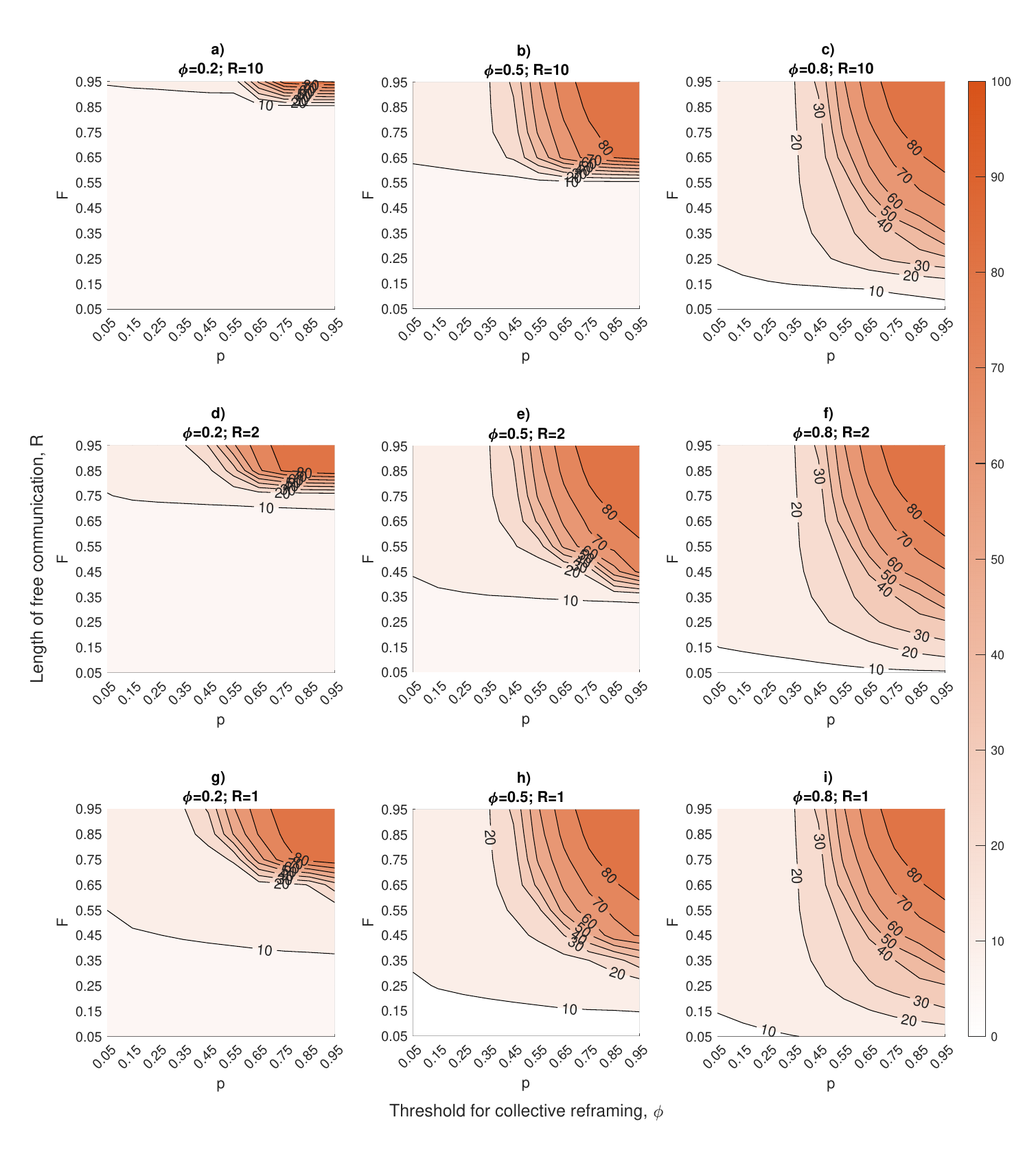}
		\label{fig:ParameterSweep_nu-R_p-F_D}} 
    \subfloat[Innovation]{
		\centering \includegraphics[width = 0.5 \textwidth]{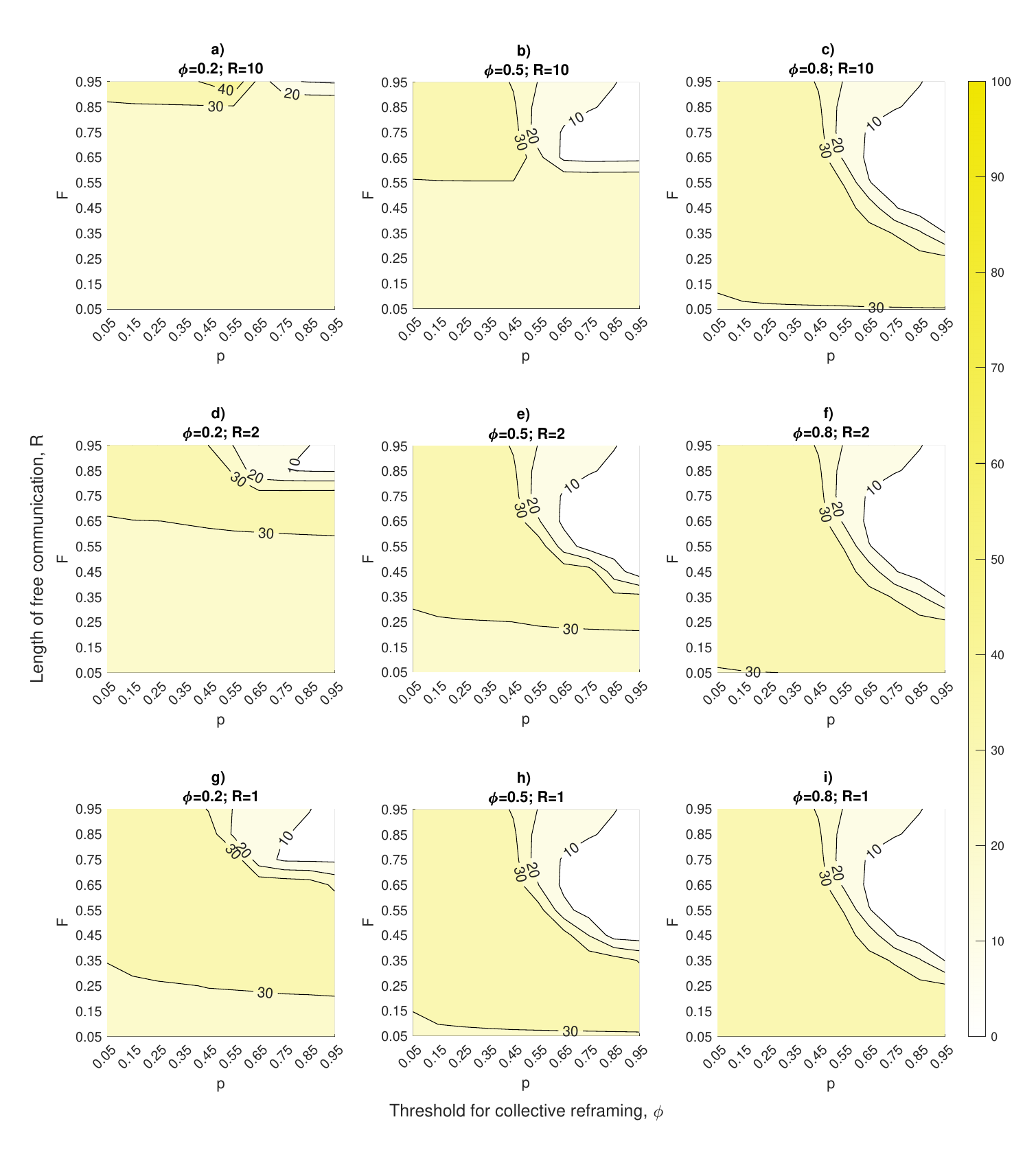}
		\label{fig:ParameterSweep_nu-R_p-F_I}}

    \subfloat[Moralisation]{
		\centering \includegraphics[width = 0.5 \textwidth]{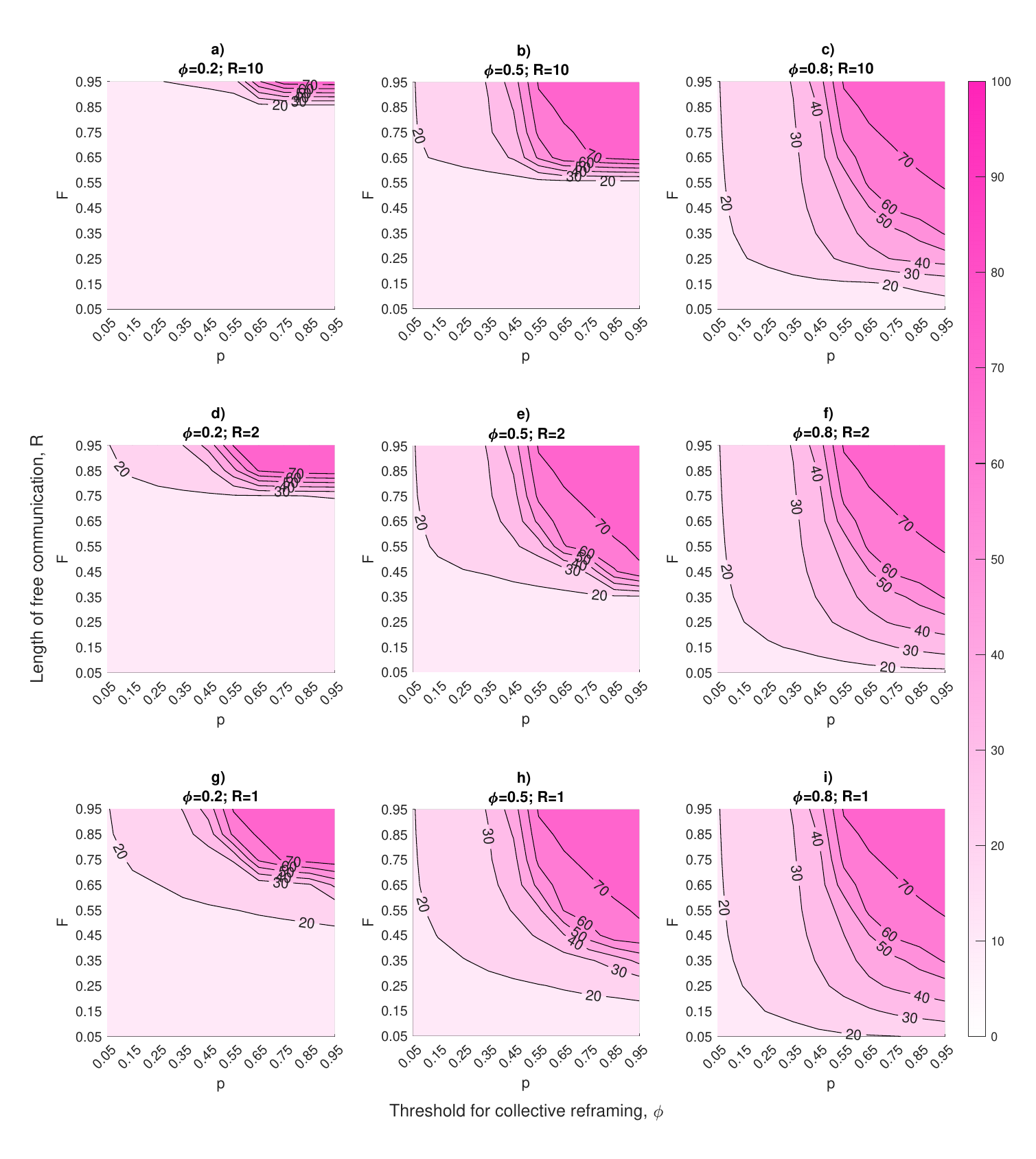}
		\label{fig:ParameterSweep_nu-R_p-F_M}} 
    \subfloat[Energisation]{
		\centering \includegraphics[width = 0.5 \textwidth]{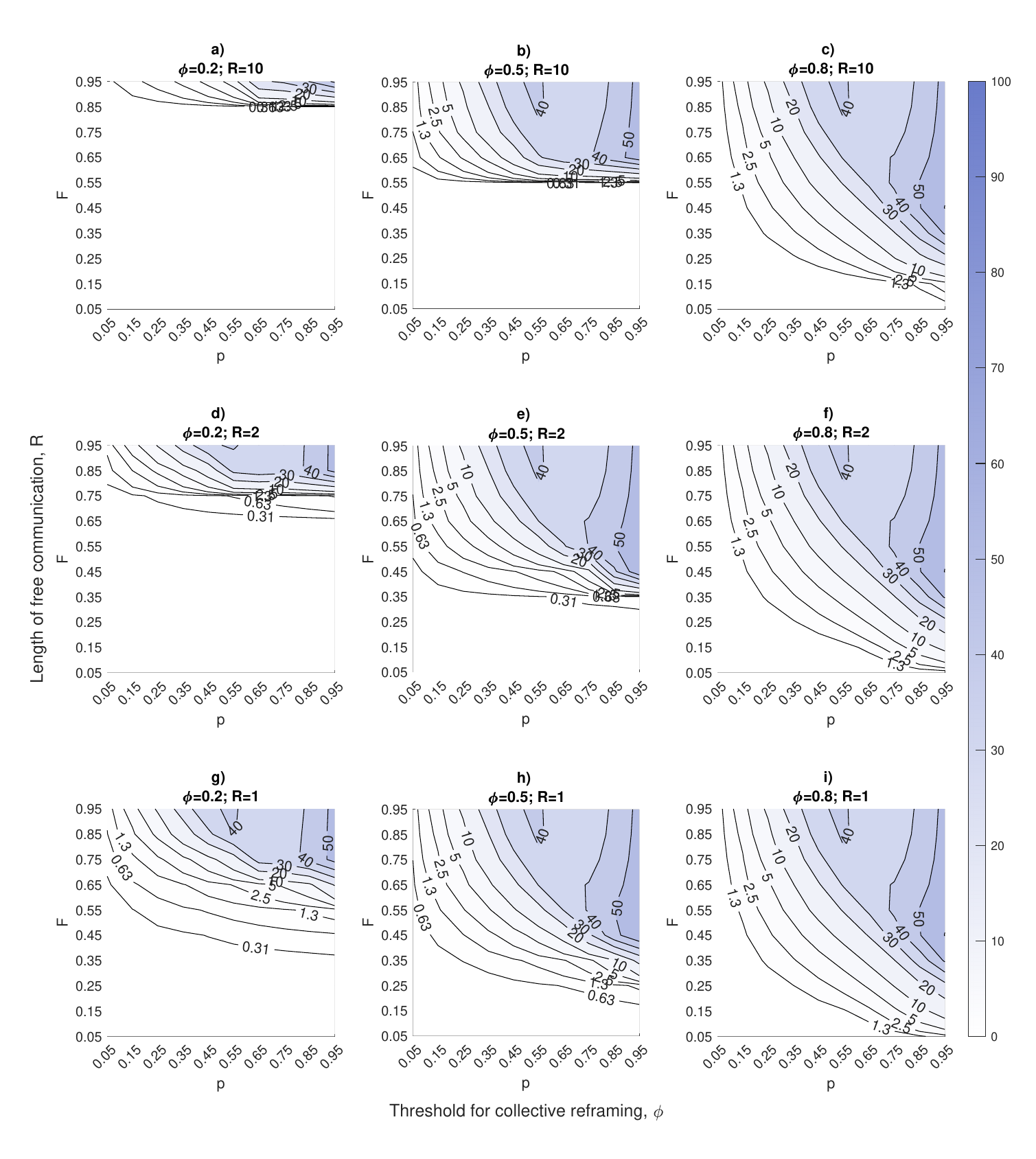}
        \label{fig:ParameterSweep_nu-R_p-F_E}}
  
    \caption{Variation for different $\phi$ (left to right) and $R$ (bottom to top) at the long-run steady state of (a-g) population fractions for the different agent types and dominant type, and (h-k) average DIME values among agents.} 
    \label{fig:ParameterSweep_nu-R_p-F}
\end{figure}

The general trend that was observed when investigating the sensitivity to $p$ and $F$ for $\phi=0.8$ and $R=10$, also holds for other combinations of $\phi$ and $R$ (see Table \ref{tab:p_F_regions}). While the `dominant agent type' refers to the type which holds the highest population fraction, there can exist other agent types which hold lower, albeit close, population fractions. For example, at low values of $p$ and $F$, both latent conventional and active innovator agents show moderate presence in the population, even though the dominant action shows only one or the other. Likewise, at medium values of $p$ and $F$, both active conventional and active innovator agents show moderate presence. Active innovators (Fig.~\ref{fig:ParameterSweep_nu-R_p-F_Inv}) hold their strongest presence when average values of innovation among agents (Fig.~\ref{fig:ParameterSweep_nu-R_p-F_I}) is at its strongest (i..e. low-medium values of $p$ and $F$), with similar contours in their plots, which matches expectation.

$\phi$ and $R$ affect the size of the individual subregions. As $\phi$ increases, the `low' $p-F$ region shrinks (as evidenced by the decreasing region of moderate presence from latent conventional agents), the `medium' $p-F$ region expands (as evidenced by the increasing region of moderate presence from active conventional agents), and the `high' $p-F$ region expands (as evidenced by the increasing region of strong presence from latent radical agents). The opposite occurs as $R$ increases, though the influence of $R$ decreases as $\phi$ increases, with changes in $R$ having marginal effect at high $\phi$.


\begin{table}
    \centering
    \begin{tabular}{c|ccc}
         & Low $p-F$  & Medium $p-F$ & High $p-F$ \\ \midrule
         Active conventional agent & Low  & Moderate  & Low \\
         Latent conventional agent & Moderate  & Low  & Marginal  \\
         Active innovator & Moderate  & Moderate  & Marginal  \\
         Latent innovator & Marginal  & Marginal  & Marginal  \\
         Active radical agent & Marginal  & Marginal  & Marginal  \\
         Latent radical agent & Low  & Low  & Strong  \\ 
         Dominant agent type & Latent Conventional, Active Innovator & Active Conventional, Active Innovator  & Latent Radical \\ \midrule
         Disidentification & Low & Moderate & High \\
         Innovation & Moderate & Moderate & Low \\
         Moralisation & Low & Moderate & High \\
         Energisation & Marginal & Low & Moderate \\ \midrule
         Effect of increasing $\phi$ & Shrinks & Expands & Expands \\
         Effect of increasing $R$ & Expands & Shrinks & Shrinks
    \end{tabular}
    \caption{Characteristics of $p-F$ subregions regarding population fractions of individual agent types, average DIME values, and effect of $\phi$ and $R$}
    \label{tab:p_F_regions}
\end{table}

\newpage
\section{Sensitivity of Population Fractions of Protestor Types and DIME Values to $\phi$ and $R$}
So far, we have presented and discussed the sensitivity of the long-run steady state agent type population fractions and average DIME values with greater emphasis on $p$ and $F$ (the axes of the prior contour plots), than $\phi$ and $R$ (3-level discretization in Fig.~\ref{fig:DomActions_Rphi} and \ref{fig:ParameterSweep_nu-R_p-F}). Here, we supplement by presenting and discussing the sensitivity with respect to $\phi$ and $R$, for fixed $p=0.85$ and $F=0.45$, of the population fractions of the six individual agent types (Fig.~\ref{fig:ParameterSweep_nu-R_Action}) and the average DIME values among agents (Fig.~\ref{fig:ParameterSweep_nu-R_DIME}). 

\begin{figure}[H]
	\centering 
    \includegraphics[width = \textwidth]{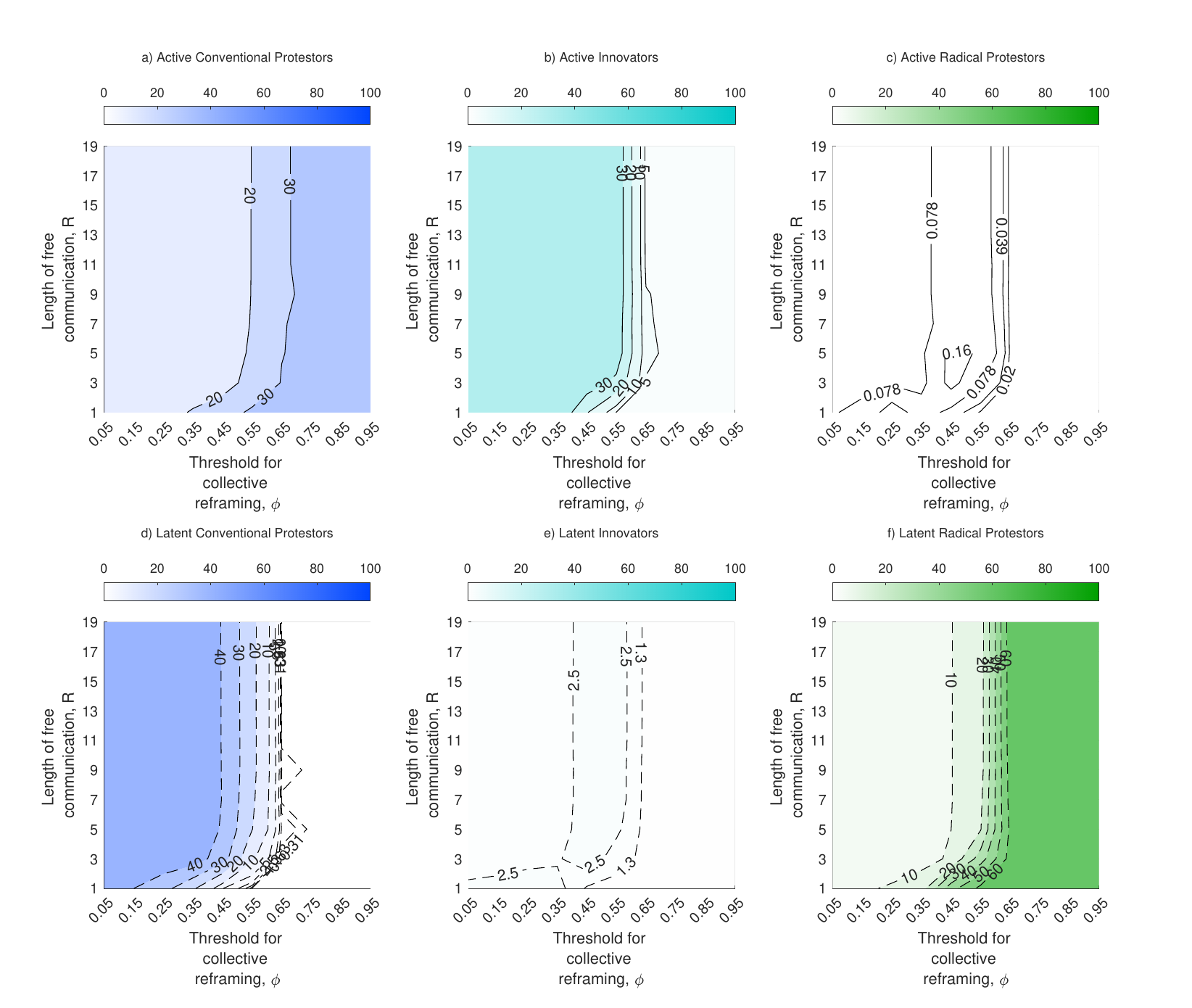}
    \caption{Long-run steady state population fractions for the six different agent types when varying $\phi$ and $R$, for fixed $p=0.85$ and $F=0.45$. Observe that changes in $R$ beyond $10$ have negligible effect.}
    \label{fig:ParameterSweep_nu-R_Action}
\end{figure}

\begin{figure}[H]
	\centering 
    \includegraphics[width = \textwidth]{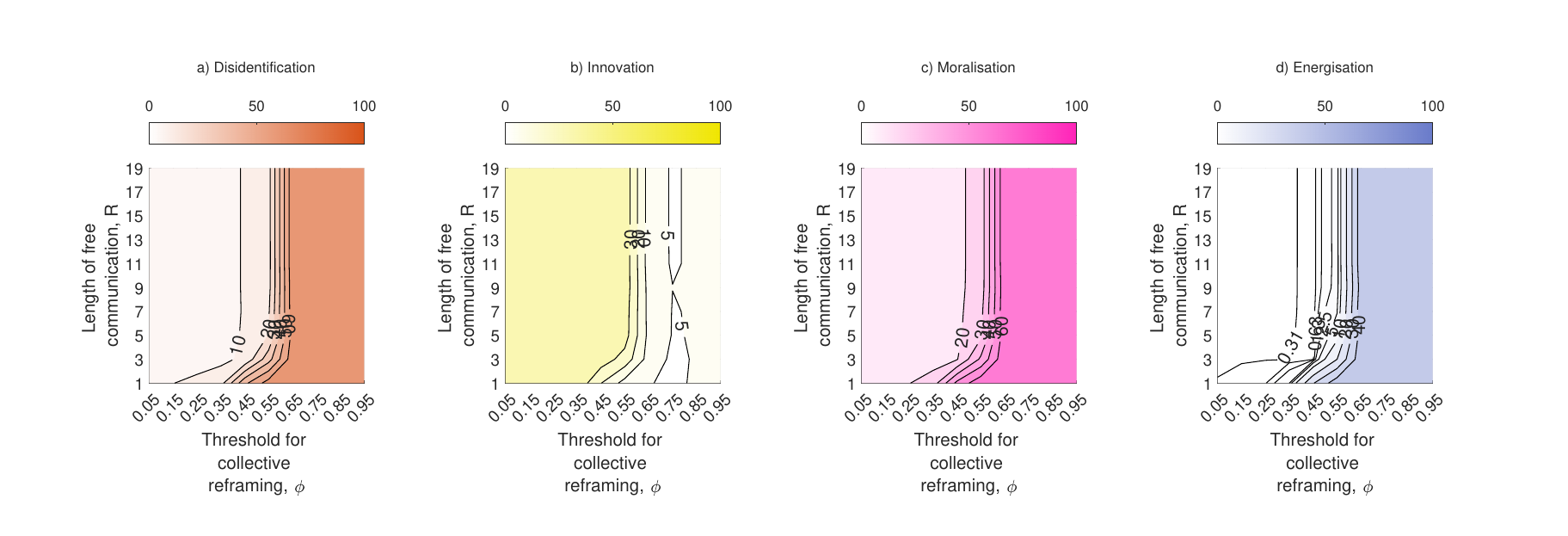}
    \caption{Long-run steady state average DIME values among agents when varying $\phi$ and $R$, for fixed $p=0.85$ and $F=0.45$. Observe that changes in $R$ beyond $10$ have negligible effect.}
    \label{fig:ParameterSweep_nu-R_DIME}
\end{figure}

As $\phi$ increases: 1) the population fractions of active conventional and latent radical generally increase; 2) the population fractions of active innovators and latent conventional generally decrease; 3) the population fractions of active radical and latent innovators remain low; 4) the average values of disidentification, moralisation, and energisation generally increase; and 5) the average values of innovation generally decrease. The sensitivity is strongest at medium values of $\phi$.

With respect to $R$, we observe marginal changes in the population fractions and DIME values beyond $R=10$. Hence, the interval $[1,10]$ captures all noticeable effects of $R$ on the steady state values, with the most substantial effects occurring in $[1, 4]$. This interval of noticeable effect further shrinks depending on the value of $\phi$; e.g. for this instance of $p=0.85$ and $F=0.45$, $R$ has marginal effect on the steady state values when $\phi \leq 0.15$ or $\phi \geq 0.65$. When $R$ does have a noticeable effect, it is opposite to that of $\phi$; e.g. latent radicalism generally increases with $\phi$ but decreases with $R$.

\clearpage
\section{Sensitivity of Population Fractions of Protester Types and DIME Values to Initial Action}
The results presented so far were borne from simulations run with the same initial conditions of all active conventional agents: $A_i(0) = 1, C_i(0) = 1, x^h_i(0) = 1 \; \forall \; i = 1, 2, \dots, n$. Here, we supplement by presenting and discussing the long-term steady state population fractions of the agent types and average DIME values in the responsive authority and intransigent authority scenarios that were discussed in the main text, for the different initial conditions listed in Table \ref{tab:initial_conditions}. Figure~\ref{fig:InitialActions} depicts barcharts of the long-run steady-state population fractions and DIME values for the different initial conditions.

\begin{table}[h!]
    \centering
    \caption{Different initial conditions analysed. The random variables were sampled uniformly from the corresponding sets.}
    \begin{tabular}{c | c c c} 
         Agent Type & $A_i(0)$ & $C_i(0)$ & $x^h_i(0)$ \\ \midrule
         All active conventional & 1 & 1 & 1 \\
         All latent conventional & 0 & 1 & 1 \\
         All active radical & 1 & 1 & -1 \\
         All latent radical & 0 & 1 & -1 \\
         Random & $x\in \{0,1\}$ & $x\in \{-1,1\}$ & $x\in \{-1,1\}$
    \end{tabular}    
    \label{tab:initial_conditions}
\end{table}

\begin{figure}[H]
    \centering
    \subfloat[Actions in Responsive Authority Scenario]{
    \centering \includegraphics[width = 0.45\textwidth]{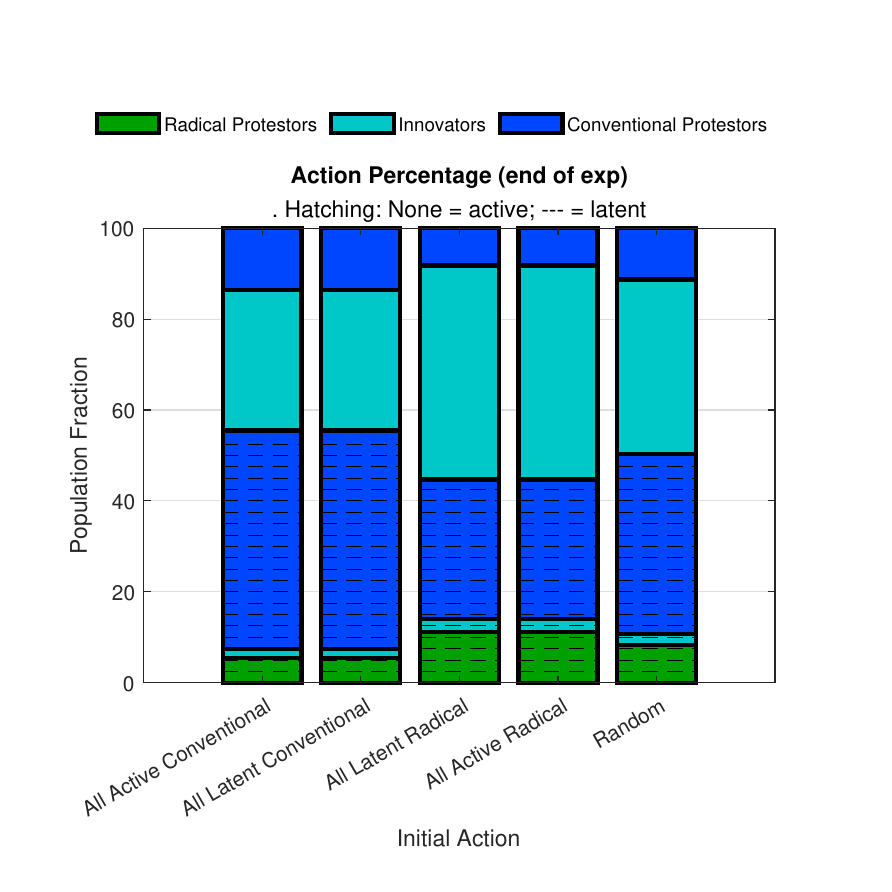}
    \label{Fig:InitialActions_CScenario_Action}
    }
    \subfloat[Actions in Intransigent Authority Scenario]{
    \centering \includegraphics[width = 0.45\textwidth]{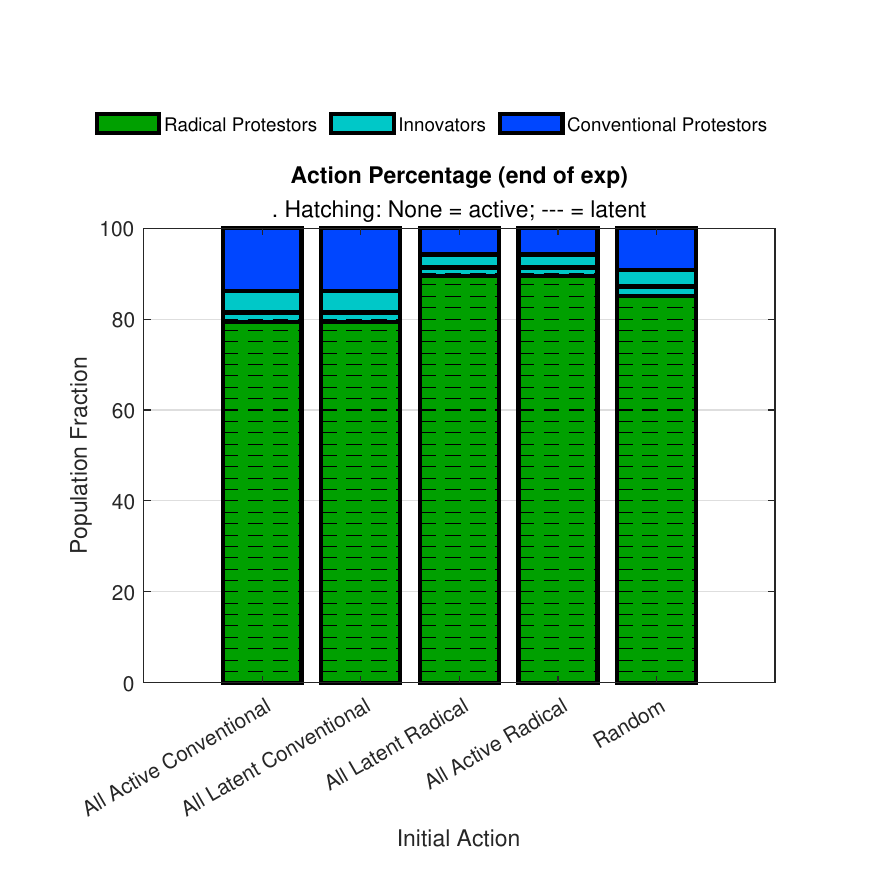}
    \label{Fig:InitialActions_LRScenario_Action}
    }
    
    \subfloat[DIME in Responsive Authority Scenario]{
    \centering \includegraphics[width = 0.45\textwidth]{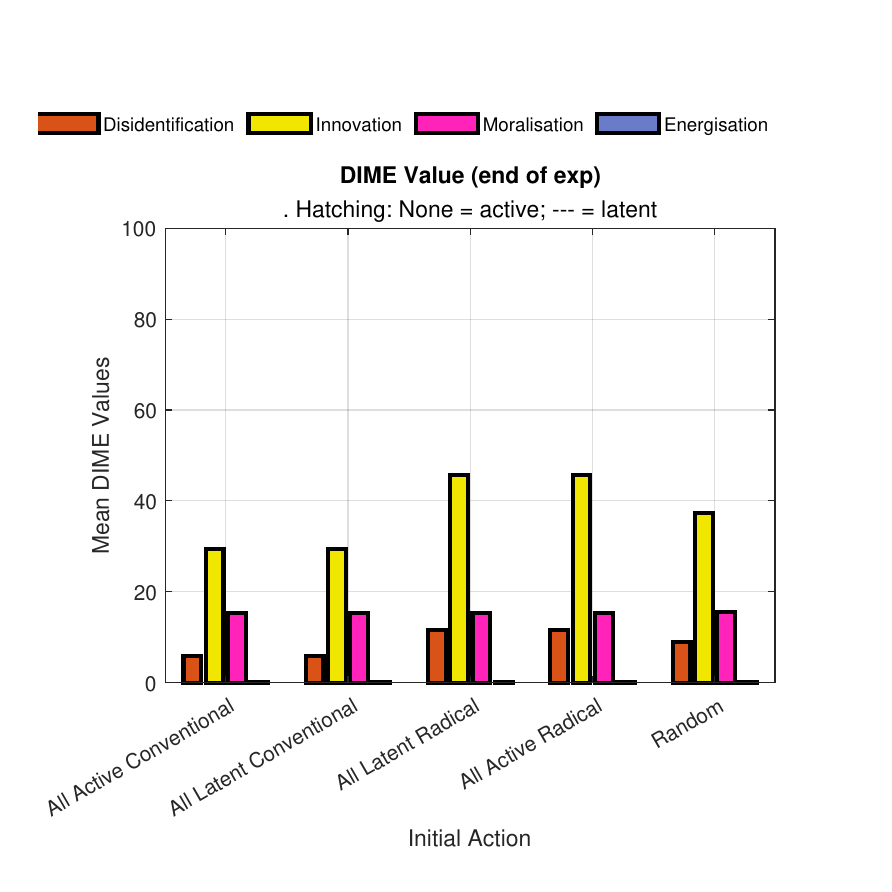}
    \label{Fig:InitialActions_CScenario_DIME}
    }
    \subfloat[DIME in Intransigent Authority Scenario]{
    \centering \includegraphics[width = 0.45\textwidth]{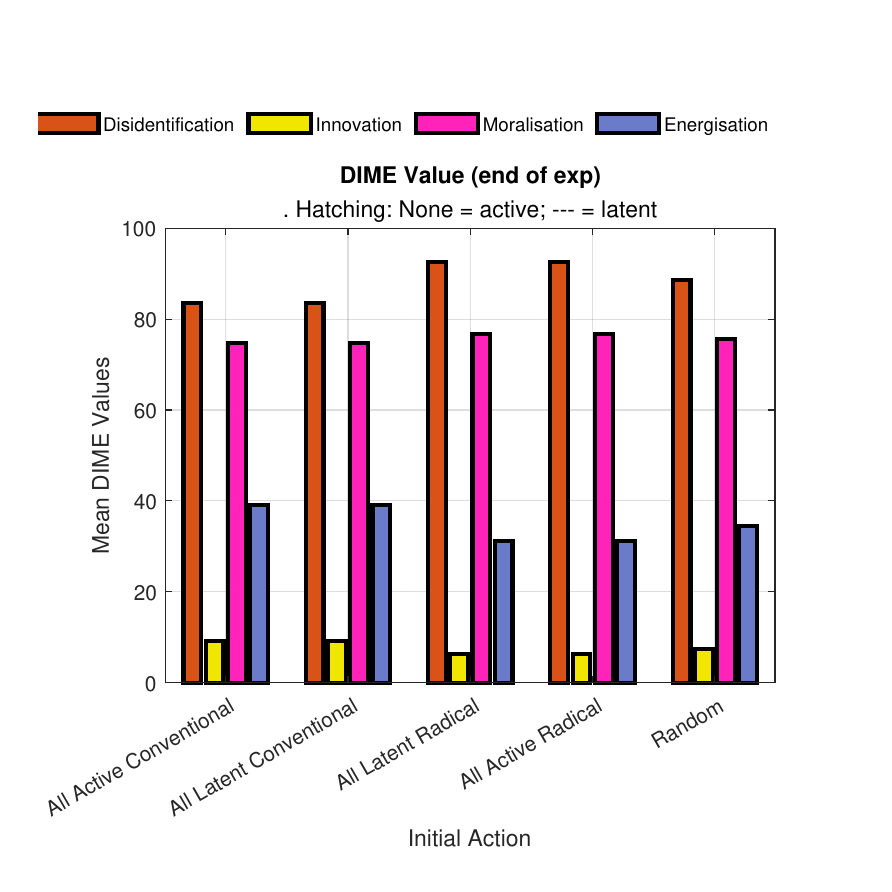}
    \label{Fig:InitialActions_LRScenario_DIME}
    }
    
    \caption{Sensitivity to initial action distribution, of long-run steady state population fractions and average DIME values among agents, in the two idealised scenarios discussed in the main text. The action bars comprise, from top to bottom: active conventional, active innovators, active radical, latent conventional, latent innovators, and latent radical. While the individual values vary, the final distribution is qualitatively insensitive to the distribution of initial actions.}
    \label{fig:InitialActions}
\end{figure}

For the population fractions in the responsive authority scenario (Fig.~\ref{Fig:InitialActions_CScenario_Action}), we observe that the `all active conventional' and `all latent conventional' initial conditions yield quantitatively similar population fractions, with latent conventional agents making up the dominant fraction. Likewise, the `all active radical' and `all latent radical' initial conditions yield quantitatively similar population fractions to each other, with active innovating agents making up the dominant fraction. The population fractions from the `random' initial conditions lie in between the two groups, roughly the average of the two (which is reasonable to expect given the equal probability of the initial condition falling under either group). Hence, the steady-state population fractions in this scenario are quantitatively insensitive to whether agents are initially active or latent. While there are quantitative differences in the population fractions based on whether agents are initially in conventional or radical orientation, the ordering of the agent types differs only by switching of the largest two types (latent conventional and active innovator) and population fractions are roughly on the same order of magnitude; implying qualitative similarity. Hence, the steady-state population fractions in this scenario are qualitatively insensitive to whether agents are initially in conventional or radical orientation. Therefore, the steady-state population fractions in this scenario are qualitatively insensitive to the initial conditions chosen.

Similar behaviour can be observed for the other results of mean DIME values in the responsive authority scenario (Fig.~\ref{Fig:InitialActions_CScenario_DIME}), and the population fractions (Fig.~\ref{Fig:InitialActions_LRScenario_Action}) and mean DIME values (Fig.~\ref{Fig:InitialActions_LRScenario_DIME}) in the intransigent authority scenario scenario. Thus, these results give us confidence to believe that the results presented in this paper are insensitive to the choice of initial conditions.

\clearpage
\section{Comparison of Network Types}
In Fig.~\ref{fig:NetworkComparison}, we present visualizations of an example of the Holme-Kim network (the network type used in our simulations) \cite{holme2002network}, and compare against the Erdős–Rényi network (random regular). The Holme-Kim network (Fig.~\ref{Fig:NetworkComparison_HK_Graph} and \ref{Fig:NetworkComparison_HK_Degree}) was generated with $N=1000, m=6, m_t = 5, $ and $N_0=13$, while the Erdős–Rényi network (Fig.~\ref{Fig:NetworkComparison_ER_Graph} and \ref{Fig:NetworkComparison_ER_Degree}) was generated with $N=1000$ and $p =0.01$ (probability of edge formation). These parameters were chosen so that the two networks have similar characteristic path length ($l$; $3.01$ for the Holme-Kim network and $3.03$ for the Erdős–Rényi network) and average degree ($<k>$; $12.00$ for the Holme-Kim network and $12.07$ for the Erdős–Rényi network). But despite these similarities, there is greater clustering in the Holme-Kim network than in the Erdős–Rényi network, as evidenced by the global clustering coefficient: $\gamma$; $0.18$ for the Holme-Kim network and $0.01$ for the Erdős–Rényi network.

\begin{figure}[H]
    \centering
    \subfloat[Graph of Holme-Kim Network]{
    \centering \includegraphics[width = 0.45\textwidth]{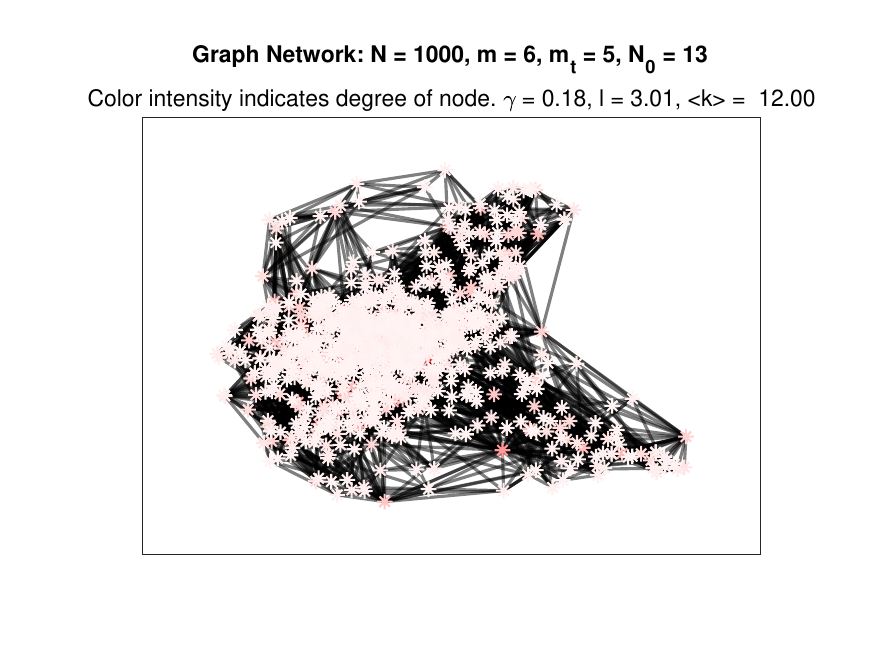}
    \label{Fig:NetworkComparison_HK_Graph}
    }
    \subfloat[Graph of Erdős–Rényi Network]{
    \centering \includegraphics[width = 0.45\textwidth]{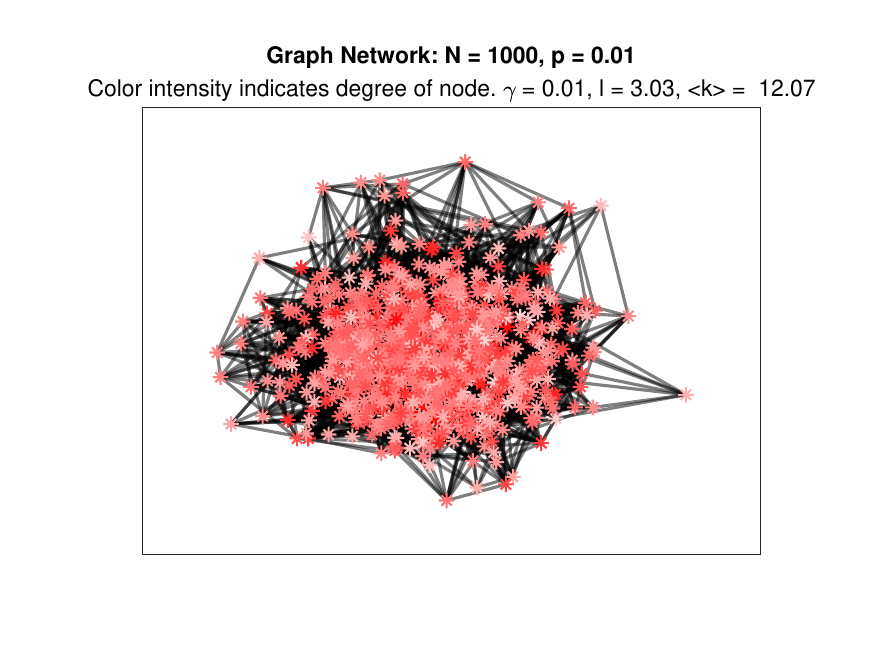}
    \label{Fig:NetworkComparison_ER_Graph}
    }
    
    \subfloat[Degree Distribution in Holme-Kim Network]{
    \centering \includegraphics[width = 0.45\textwidth]{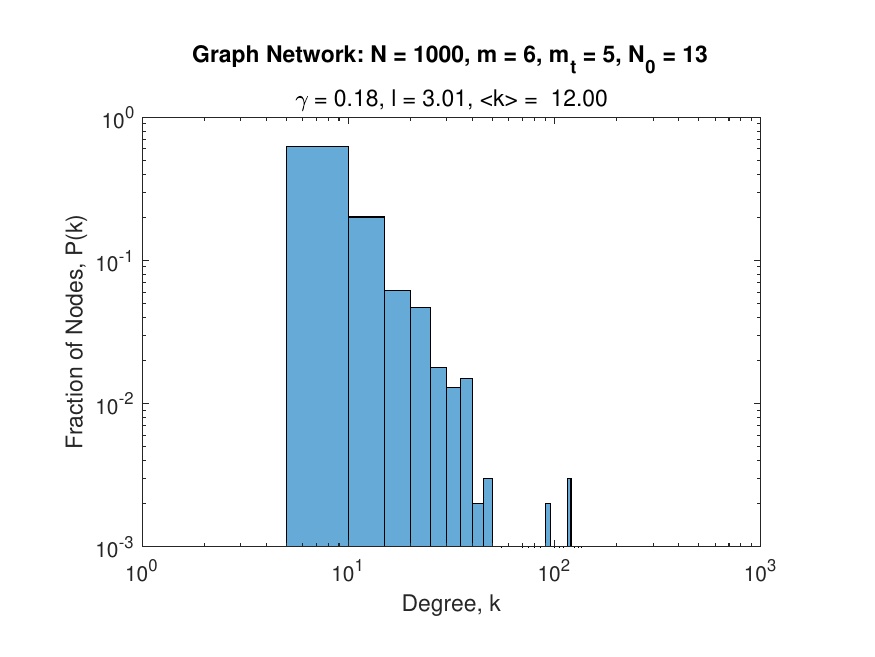}
    \label{Fig:NetworkComparison_HK_Degree}
    }
    \subfloat[Degree Distribution in Erdős–Rényi Network]{
    \centering \includegraphics[width = 0.45\textwidth]{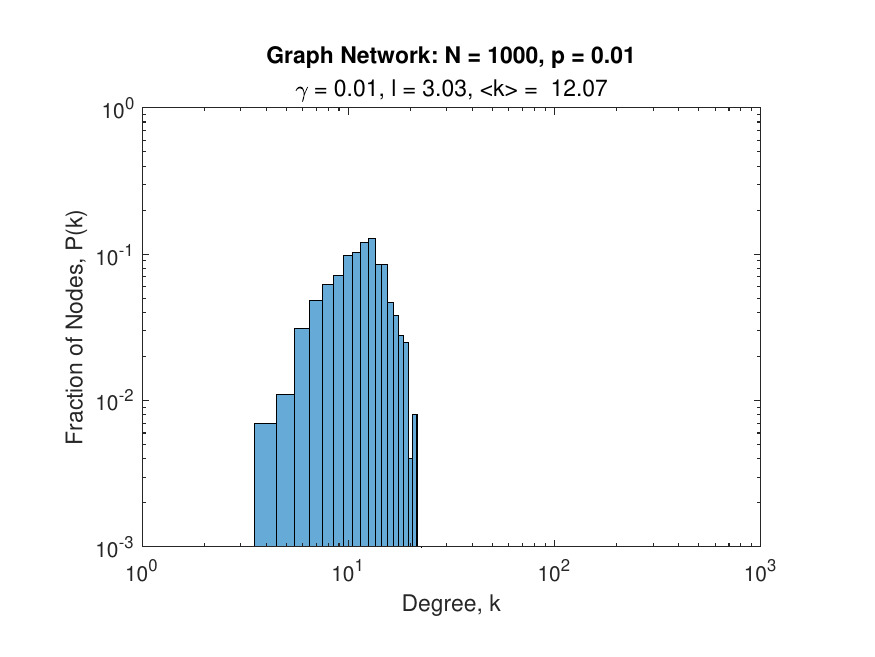}
    \label{Fig:NetworkComparison_ER_Degree}
    }
    
    \caption{Visualisation of graphs and degree distributions for examples of (a,c) Holme-Kim network \cite{holme2002network} used in our model, and (b,d) Erdős–Rényi network for comparison. The parameters of the Erdős–Rényi network were chosen so that it has similar characteristics (size, $N$; path length, $l$; average degree, $<k>$) as the Holme-Kim network example. Intensity of node color indicates node degree. Note the power-law degree distribution and the higher clustering coefficient ($\gamma$) for the Holme-Kim network.} \label{fig:NetworkComparison}
\end{figure}

Figures~\ref{Fig:NetworkComparison_HK_Degree} and \ref{Fig:NetworkComparison_ER_Degree} show the degree distributions for the two networks, from which we can observe that the Holme-Kim network has a power-law (scale-free) degree distribution, in contrast to the Erdős–Rényi network which has a normal distribution. Therefore, Holme-Kim networks are suitable for systems where power-law degree distribution with high clustering are expected.
\fi

\newpage

\end{document}